\documentclass[%
 reprint,
 amsmath,amssymb,
 aps,
]{revtex4-2}

\usepackage{hyperref}
\usepackage{graphicx}
\usepackage{dcolumn}
\usepackage{bm}

\usepackage{xcolor}
\definecolor{cream}{RGB}{222,217,201}

\begin{document}

\preprint{APS/123-QED}

\title{Chemotaxis-Driven Instabilities Govern Size, Shape and Migration Efficiency of Multicellular Clusters}

\author{Monika Sanoria\textit{$^{a}$}}
\email{msanoria@ucmerced.edu} 
\author{Gema Malet-Engra\textit{$^{b}$}}
\author{Giorgio Scita\textit{$^{bc}$}}
\author{Nir Gov\textit{$^{de}$}}
\email{nir.gov@weizmann.ac.il} 
\author{Ajay Gopinathan\textit{$^{af}$}}
\email{agopinathan@ucmerced.edu}

\affiliation{\textit{$^{a}$~Center for Cellular and Biomolecular Machines, University of California Merced, CA, 95343, USA. }}
\affiliation{\textit{$^{b}$~IFOM ETS - The AIRC Institute of Molecular Oncology, Via Adamello 16, Milan, 20139, Italy.}}
\affiliation{\textit{$^{c}$~The Department of Oncology and Hemato-Oncology at the University of Milan, Via Festa del Perdono 7, 20122 Milano, Italy.}}
\affiliation{\textit{$^{d}$~Department of Chemical and Biological Physics, Weizmann Institute of Science, 7610001 Rehovot, Israel.}}
\affiliation{\textit{$^{e}$~Department of Physiology, Development and Neuroscience, University of Cambridge, Cambridge CB2 3DY, United Kingdom.}}

\affiliation{\textit{$^{f}$~Department of Physics, University of California Merced, CA, 95343, USA. }}

\date{\today}

\begin{abstract}
The collective chemotaxis of multicellular clusters is an important phenomenon in various physiological contexts, ranging from embryonic development to cancer metastasis. Such clusters often display interesting shape dynamics and instabilities, but their physical origin, functional benefits, and role in overall chemotactic migration remain unclear. Here, we combine computational modeling and experimental observations of malignant lymphocyte cluster migration {\it in vitro} to understand how these dynamics arise from an interplay of chemotactic response and inter-cellular interactions. Our cell-based computational model incorporates active propulsion of cells, contact inhibition of locomotion, chemoattractant response, as well as alignment, adhesive, and exclusion interactions between cells. We find that clusters remain fluid and maintain cohesive forward migration in low chemoattractant gradients. However, above a threshold gradient, clusters display an instability driven by local cluster-shape dependent velocity differentials that causes them to elongate perpendicular to the gradient and eventually break apart. Comparison with our {\it in vitro} data shows the predicted transition to the cluster instability regime with increased gradient, as well as quantitative agreement with key features such as cluster aspect ratio, orientation, and breaking frequency. This instability naturally limits the size of multicellular aggregates, and, in addition, clusters in the instability regime display optimal forward migration speeds, suggesting functional implications {\it in vivo}. Our work provides valuable insights into generic instabilities of chemotactic clusters, elucidates physical factors that could contribute to metastatic spreading, and can be extended to other living or synthetic systems of active clusters.
\end{abstract}

\maketitle

\section{Introduction}

Collective migration is a widely observed phenomenon across diverse biological systems, ranging from bird flocks and fish schools to bacterial colonies and migrating cell clusters \cite{vicsek2012collective, shellard2020rules, mcmillen2024collective, Gompper_2025, dreyer2025comparing, feinerman2018physics,Elgeti2015, tunstrom2013collective, Gopinathan2019, shi_collective_2025,Bhattacharjee2022, hakim2017collective}. Despite differences in scale and biological organization, these systems exhibit emergent behaviors driven by local interactions among individual agents. For instance, bird flocks and fish schools rely on visual and hydrodynamic cues \cite{lopez2012behavioural, toner2005hydrodynamics, Hemelrijk2012}, while bacterial colonies and cell clusters depend on chemical signaling, adhesion forces, and mechanical constraints \cite{Bhattacharjee2022, Beer2019, friedl2009collective, camleyPhysicalModelsCollective2017, copenhagen2018frustration, malet-engraCollectiveCellMotility2015}. A unifying feature of these systems is their ability to self-organize and adapt to environmental changes without central control.

This feature is evident in collective cell migration, where groups of cells move in a coordinated manner through mechanical and biochemical interactions \cite{rorth2009, spatarelu2019biomechanics, colizzi_evolution_2020, camleyPhysicalModelsCollective2017, amintas2020circulating, Gopinathan2019, shi_collective_2025, varennes2016collective,serra_dynamic_2020,tan2022odd,lin_dynamic_2019,dowdell_competition_2023}. Unlike single-cell migration, which is primarily governed by intracellular signaling and cytoskeletal dynamics \cite{thuroff_bridging_2019, Rappel2017}, collective migration is also controlled by intercellular adhesion, mechanical coupling, and cooperative response to external cues \cite{Gopinathan2019, camleyPhysicalModelsCollective2017, copenhagen2018frustration, malet-engraCollectiveCellMotility2015}. These properties enable cell clusters to maintain cohesion while navigating complex microenvironments \cite{cristini2005morphologic}, making collective migration essential for various physiological and pathological processes, including tissue formation, regeneration, immune response, and cancer metastasis \cite{friedl2009collective, camleyPhysicalModelsCollective2017, friedl2012classifying}. Understanding the principles governing this behavior provides crucial insights into both normal development and disease progression, particularly in cases where collective migration facilitates metastasis or wound healing.

Among various forms of collective migration, ranging from cohesive sheets \cite{poujade2007collective1} to invasive strands and expanding tumors \cite{cristini2005morphologic}, migrating cell clusters \cite{Camley2018} represent a crucial system for studying mechanics, guidance, and emergent behavior due to their confined size and responsive cohesive structure. Although there has been work on shape instabilities in non-migrating systems such as tumors \cite{cristini2005morphologic, Alert2020}, as well as the influence of substrate properties \cite{sunyer2016collective} and confinement \cite{Ilina2020,wang_confinement_2025} on their behavior, the dynamics of actively migrating clusters remains less understood.
Migrating clusters respond to various directional cues \cite{haeger2015collective}: electric fields guide collective movement in galvanotaxis \cite{copos2025galvanotactic}, substrate stiffness gradients direct migration through durotaxis \cite{pi2022collective}, and chemical gradients influence movement through chemotaxis. 
Cellular response to such cues can depend on positioning within clusters \cite{camley2016emergent} leading to internal rearrangements in fluid clusters \cite{malet-engraCollectiveCellMotility2015, copenhagen2018frustration}. Chemical gradients can therefore induce asymmetric migration, potentially leading to instability, deformation, or fragmentation \cite{malet-engraCollectiveCellMotility2015}; however, their precise contribution to cluster dynamics and migration efficiency is not well characterized. 

\begin{figure}
  \centering
  \includegraphics[width=\linewidth]{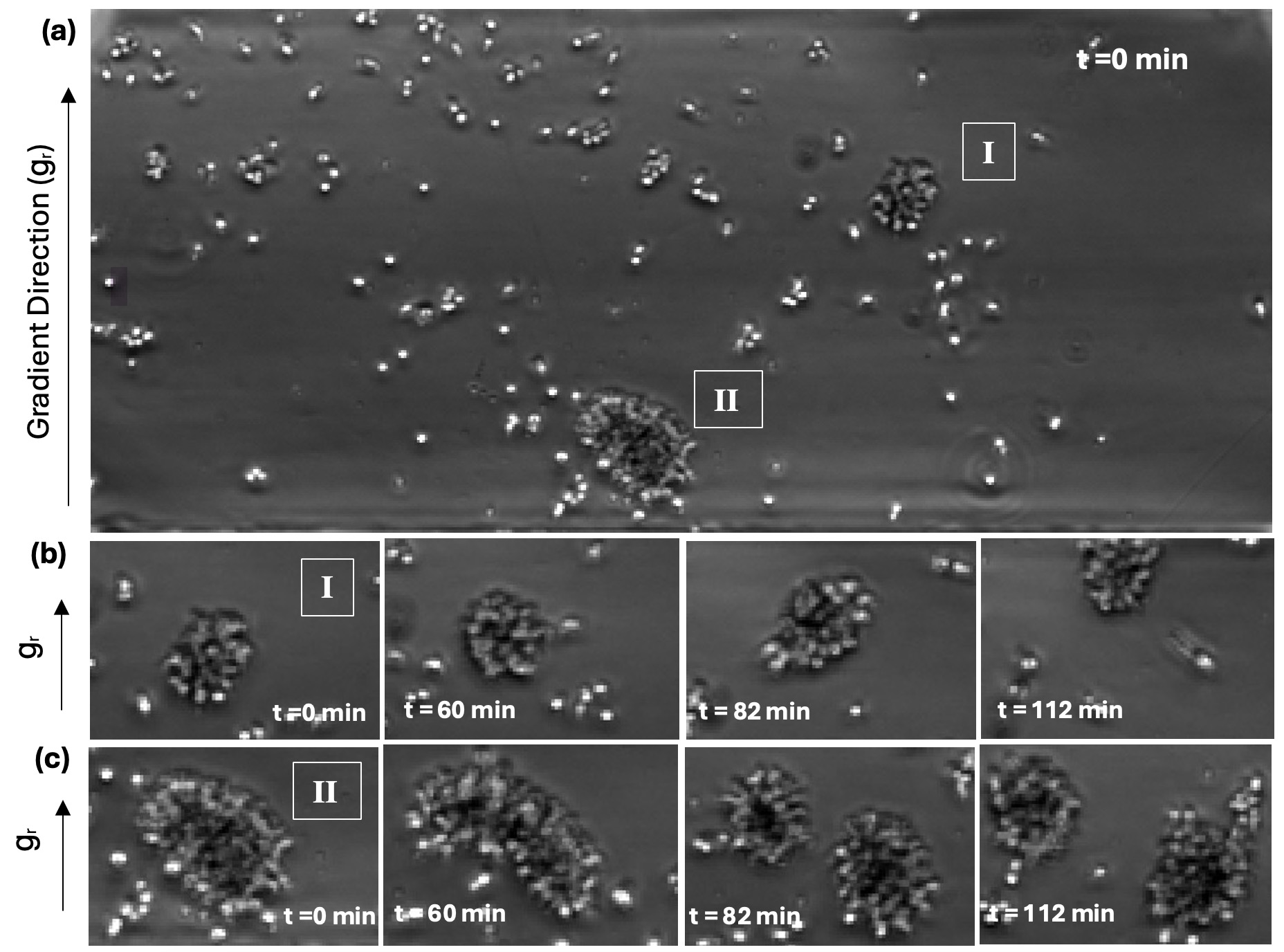}
  \caption{(a) Initial snapshot of the experiment, showing multiple clusters, including two prominent ones of different sizes. (b) Time series of a smaller cluster (I) that remains stable and cohesive over time. (c) Time series of a larger cluster (II) that undergoes deformation, elongation, and eventual fragmentation. These contrasting behaviors suggest a competition between stabilizing and destabilizing factors, raising the question of what governs cluster integrity versus breakup.}
  \label{FIG.main1}
\end{figure}

Here, we analyze experimental data on malignant B lymphocytes in a chemotaxis assay (in a flat, two-dimensional geometry, see Supplementary Video M1) exposed to a gradient of the homeostatic chemokine CCL19 \cite{malet-engraCollectiveCellMotility2015}, which acts as a chemoattractant. Cluster cohesion is maintained through LFA-1/ICAM-1 interactions (integrin/receptor binding) between cells, which are known to be mediated by divalent cations such as Ca$^{2+}$ or Mn$^{2+}$. Under gradients of CCL19, the clusters in the experiments exhibited shape instabilities, elongating and eventually fragmenting (see Fig~\ref{FIG.main1}, and Supplementary Video M2). Larger clusters broke apart more readily, while smaller clusters exhibited a higher threshold for fragmentation. These observations suggest that the instability is influenced not only by chemoattractant gradients but also by the size and mechanical properties of the clusters themselves. 


To understand the fundamental principles underlying the shape dynamics and instabilities of multicellular clusters during chemotactic migration, we used a computational model, incorporating active propulsion of cells, contact inhibition of locomotion, chemoattractant response, as well as alignment, adhesive, and exclusion interactions between chemoattractant cells. 
Our findings reveal that at low chemoattractant gradients, below a critical threshold, clusters remain fluid and maintain cohesive forward migration. Above this threshold, the clusters become unstable, elongating perpendicular to their direction of motion and eventually fragmenting. This instability is driven by a feedback mechanism due to differences in the response of cells depending on the local cluster shape. Furthermore, we showed that this instability is also a function of cluster size, with larger clusters exhibiting higher susceptibility to deformation and fragmentation. At even higher gradients, the chemotactic response leads to cluster solidification, significantly reducing forward migration. Clusters in the instability regime have their sizes naturally regulated and show optimal forward migration speeds, suggesting functional benefits {\it in vivo}. Our theoretical predictions align closely with our {\it in vitro} observations of migrating clusters, providing insights into collective cell behavior with implications for tissue development and disease progression as well as active matter clusters in general.

\section{Model}
\begin{figure*}[htbp]
 \centering
\includegraphics[width=0.98\linewidth]{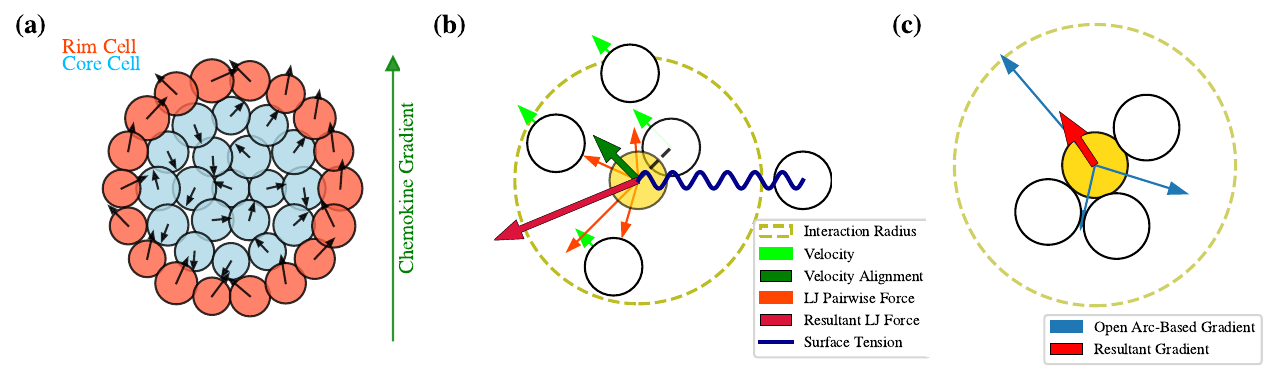}
\caption{\textbf{Schematic Representation of the Model} 
(a) {\bf Cell Cluster with Directed Forces}: A cluster of rim cells (red particles) and core cells (blue particles) is shown with arrows representing velocity vectors. The cluster responds to a chemical gradient, influencing the alignment and movement of individual cells. 
(b) {\bf Particle Interactions}: This panel illustrates the key forces experienced by a cell (yellow-highlighted particle) due to its neighbors within the interaction radius (dotted olive-colored boundary). Lennard-Jones ($\vec{LJ}_{ij}$) pairwise forces (red arrows) generate a resultant $\vec{LJ}_{i}$ force (dark red, Eq.~\ref{LJ}), that influences cell positioning. Additional surface-tension-like forces (blue springs, Eq.~\ref{s}) are used to maintain cluster integrity. Each cell's velocity vector is shown in light green (Eq.~\ref{v}) and the mean velocity in dark green (Eq.~\ref{VV}).
(c) {\bf Gradient Sensing and Directed Migration}: The response of a cell (yellow-highlighted particle) to an external chemokine gradient is based on its exposed boundary segments (open arcs). Direction and magnitude of gradient vector due to each open arc of the yellow particle is represented by blue vectors, $\vec{f}_j$ (in Eq.~\ref{g}) and the resultant gradient vector, $\vec{g}_i$ in Eq.~\ref{g}), is shown by a red arrow. 
}
\label{FIG.main2}
\end{figure*}

We employed an agent-based model \cite{copenhagen2018frustration} to simulate cell clusters in a two-dimensional (2D) environment (see Fig.~\ref{FIG.main2} for a schematic representation). Each cell cluster consists of \(N\) particles, with a distribution of radii with polydispersity of approximately 10\% to account for natural variations in particle size and to prevent artificial ordering effects. Each particle is subject to its own active self-propulsion, interacts with other cells through exclusion and cohesion interactions and is influenced by an external chemoattractant gradient (applied in the \(y\)-direction). 

The dynamical equation of motion \cite{copenhagen2018frustration} for these particles is given by:
\begin{equation}
\vec{v}_i(t) = \vec{F}_i^p + \vec{F}_i^c + \eta_i,
\label{v}
\end{equation}
where \( \vec{F}_i^p\), \( \vec{F}_i^c\) and \( \eta_i\) are the propulsive force, cohesive force and noise experienced by the particle and where we assume overdamped dynamics. The total force acting on a particle is thus composed of three main components:

1. {\it Active Self-propulsion} (\( \vec{F}_i^p = p_i \hat{n}_i \)):
 The active speed (\(p_i\)) or propulsion of particles is governed by contact inhibition of locomotion (CIL) \cite{copenhagen2018frustration}, implying that cell propulsion is smaller when cells are confined by neighbors, with particles at the rim of the cluster moving faster than those in the core of the cluster. This is a result of cells at the surface having more open space available to form protrusions than those at the center, leading to higher propulsion forces. The propulsion of a cell is taken to be \cite{copenhagen2018frustration}:

 \begin{equation}
   p_i = p_{\text{core}} - \frac{3}{7}(p_{\text{rim}} - p_{\text{core}})(z_i - 7),
    \label{p}
  \end{equation}
  where \(p_i\), \(p_{\text{core}}\), and \(p_{\text{rim}}\) represent the propulsion of cell \(i\), core cells, and rim cells, respectively, and \(z_i\) is the number of neighbors of cell \(i\). Cells at the surface are called rim cells, while those inside the cluster, excluding the surface, are called core cells (Fig.~\ref{FIG.main2}(a)).

  Cells tend to align their velocity with the mean orientation of their neighboring cells and in the direction of their response to the chemical gradient, such that the instantaneous direction of active propulsion is given by the unit vector:

  \begin{equation}
    \hat{n}_i = \frac{\hat{v}_i(t - \Delta t) + \alpha \hat{V}_i + \beta \vec{g}_i \chi(y_i)}{\left|\hat{v}_i(t - \Delta t) + \alpha \hat{V}_i + \beta \vec{g}_i \chi(y_i)\right|}
    \label{n}
  \end{equation}
  where \(\hat{v}_i(t - \Delta t)\) is the unit velocity of cell \(i\) at the previous time step, \(\alpha\) is the strength of the alignment interaction, and \(\hat{V}\) is the unit vector of the mean orientation of nearest-neighboring cells, (for more details, see Eq.~\ref{firstCutOff} of Appendix~\ref{SI.2}), defined as (Fig.~\ref{FIG.main2}(b)):

  \begin{equation}
    \hat{V}_i = \frac{\sum_{\text{n.n.}} \vec{v}_j}{|\sum_{\text{n.n.}} \vec{v}_j|}.
    \label{VV}
  \end{equation}

  The cell's overall response to the chemoattractant is governed by the response to the local concentration of chemoattractant at the location of the cell, \(\beta\chi(y_i)\), and the local gradient vector, $\vec{g}_i$. The local gradient vector is defined as (see Fig.~\ref{FIG.main2}(c)):
  \begin{equation}
    \vec{g}_i = \sum_j \vec{f}_j,
    \label{g}
  \end{equation} 
  where \(\vec{f}_j\) is a vector pointing in the direction bisecting the angle subtended by the centers of the neighbor pair at the center of cell \(i\) (see Fig.~\ref{FIG.main2}(c)). Its magnitude corresponds to the open arc length between the two neighboring cells, that is, the arc such that rays from the center of cell $i$ passing through the arc will pass between the neighboring cells. The summation runs over all adjacent neighbor pairs of the cell. The local gradient vector thus reflects the direction and magnitude of the most exposed area of the cell boundary.

The response to the local concentration, \( \beta\chi(y_i)\), is assumed to be proportional to the bound fraction of chemoattractant (chemokine) receptors, \( \chi(y_i)\), and is taken to be
\begin{equation}
 \ \beta \chi(y_i)= \beta ~ \frac{c(y_i)}{c(y_i) + c_0},
\label{cy_sat}
\end{equation}
where we assume Langmuir adsorption and $c_0$ and $\beta$ are constants. 
The local concentration \(c(y_i)\) of each particle is given by
\begin{equation}
 c(y_i) = ((y_i - y_{cm}) + y_0) \cdot g_r,
\label{cy}
\end{equation}
where, $y_i$, and $y_{cm}$ are y-positions of the $i$th particle and CM of the cluster respectively and \(g_r\) represents the slope of the chemokine concentration gradient. We ensure $y_0 > (y_i - y_{cm}) $ for our simulations to prevent artificially negative values.

2. {\it Cohesive Forces} (\( \vec{F}_i^c = \vec{LJ}_i \)):
  This accounts for both attraction and repulsion, balancing cohesion and particle avoidance, and is given by:
  \begin{equation}
    \vec{LJ}_i = 24 \epsilon \sum_{j \neq i}^{n_b} \left[ 2 \left( \frac{\sigma_{ij}}{r_{ij}} \right)^{13} - \left( \frac{\sigma_{ij}}{r_{ij}} \right)^7 \right] \hat{r}_{ij},
    \label{LJ}
  \end{equation}
  where \(\epsilon\) is the depth of the potential well, \(\sigma_{ij}\) denotes the characteristic size (diameter) of the particle pair, defined as \(\sigma_{ij} = \frac{\sigma_i + \sigma_j}{2}\), where \(\sigma_i\) and \(\sigma_j\) are the diameters of particles \(i\) and \(j\). \(r_{ij}\) is the distance between particles \(i\) and \(j\), and \(\hat{r}_{ij}\) is the unit vector pointing from particle \(i\) to particle \(j\) (Fig.~\ref{FIG.main2}(b)). The Lennard-Jones force is only applied to particle pairs identified within the first neighbor shell, as defined by Eq.~\ref{firstCutOff} (for more details, see Appendix~\ref{SI.2}).

To study clusters of a fixed size, we minimize slow evaporation of single cells from the cluster by maintaining cohesion between second-nearest neighbors (s.n.n.) (for more details see Appendix \ref{SI.2}) via a spring-like force that acts between cells when two conditions are met: (i) the pair of cells are separated by more than the first neighbor interaction range (Eq.~\ref{firstCutOff}), and (ii) no intermediate cell exists between cell \(i\) and cell \(j\), represented by \(\omega_j = 1\). If an intervening cell is present, \(\omega_j = 0\).

The force acting on cell \(i\) is given by
\begin{equation}
  \vec{S}_i = -k_s\sum_{j}^{\text{s.n.n.}} \omega_j \vec{d}_{ij},
  \label{s}
\end{equation}
where \(k_s\) is the spring constant, \(\vec{d}_{ij}\) is the vector connecting the center of cell \(i\) to the center of cell \(j\) (\(\vec{d}_{ij} = \vec{r}_j - \vec{r}_i\)) (Fig.~\ref{FIG.main2}(b)).
This interaction occurs rarely, mainly involves rim cells that temporarily lack immediate neighbors and does not affect the cluster dynamics. 

3. {\it Stochastic Noise Vector} (\(\eta_i\)):
 This introduces random fluctuations to mimic biological variability, and is taken to be:
\begin{equation}
  \eta_i = \sqrt{2 D \Delta t} \, \xi_i,
  \label{eta}
\end{equation}
 where \(D\) is the noise strength (diffusion coefficient), \(\Delta t\) is the time step, and \(\xi_i\) is a Gaussian random vector with zero mean and unit variance.

Simulations were conducted in a setup analogous to a treadmill, where the center of the cellular cluster was fixed in space. This approach allowed us to study the cluster dynamics while the external concentration and gradient do not change over time. 
The system was simulated for particle counts ranging from \(N = 10\) to \(150\), with the following parameter values: \(\alpha = 6\), \(\epsilon = 18\), \(k_s = 0.1\), \(p_{core} = 1.5\), \(p_{rim} = 8\), \(|\eta| = 1.5\), \(c_0 = 40\), \(\beta = 80\), \(\Delta t = 0.01\). For all the simulations $y_0=250$, until mentioned. These values, identified in \cite{copenhagen2018frustration}, reproduce the observed running and rotating phases of cellular clusters during collective migration in low chemotactic gradients.
Simulations were performed up to 4000 time steps (each time step = 30s, estimated from comparing cluster speed and cell size (see Appendix~\ref{SI.1})), beyond which we did not observe any significant changes.

\begin{figure*}[htbp]
  \centering
  \includegraphics[width=0.98\linewidth]{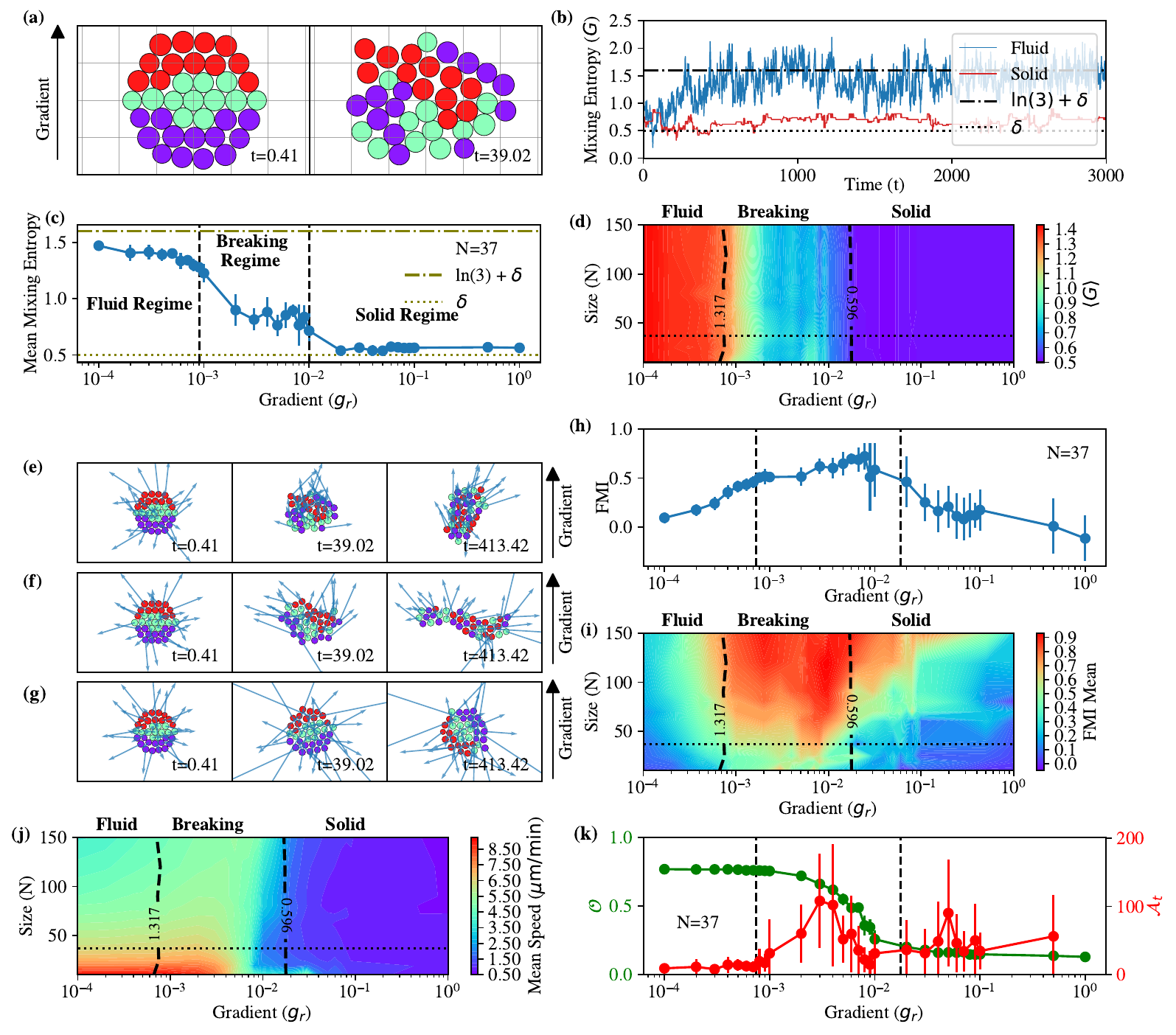}
  \caption{\textbf{\bf Chemotactic Efficiency and Behavioral Transitions of Migrating Clusters}
  (a) Representative snapshots of a migrating cluster ($N=37$) at earlier and later times illustrating mixing dynamics in the presence of a gradient. Particles are initially color-coded to track mixing over time. Binning regions of fixed size are indicated by thin gray lines.
  (b) Mixing entropy over time for ($N=37$), comparing fluid ($g_r =0.0005$) and solid ($g_r =0.2$) cases. Baselines for $\delta$ and $ln(3)+\delta$ are included.
  (c) Mixing entropy as a function of gradient, showing transitions between fluid, breaking, and solid cluster regimes. Vertical dashed lines mark critical transitions at 10\% of maximum and minimum entropy for \(N = 37\). All these quantities are measured up to the point of cluster breakup, unless specified otherwise.
  (d) Heatmap of mixing entropy as a function of size and gradient. Contour lines indicate 10\% of maximum and minimum entropy, with \(N = 37\) marked by a horizontal dashed line.
  Representative particles are colored as explained in (a), with velocity vectors shown 
  (e) Fluid clusters with complete mixing for low gradients. 
  (f) Partial mixing cluster with structural elongation prior to breakup for intermediate gradients. 
  (g) Solid clusters with no mixing for high gradients.
  (h) FMI vs. gradient, showing a non-monotonic trend. Dotted black lines mark entropy-based transitions, as in (d).
  (i) FMI heatmap depicting variations in migratory efficiency across gradients and sizes. A horizontal dashed line marks \(N = 37\), with entropy contour lines as in (d).
  (j) Heatmap of mean cluster speed as a function of gradient and size, showing increased motility at low and intermediate gradients but reduced motility at high gradients. A horizontal dashed line marks \(N = 37\), with entropy contours as in (d).
  (k) Cluster dynamics across gradients, showing low angular momentum (\(\mathcal{A}_t\), Eq.~\ref{a_max}) but high polarization (\(\mathcal{O}\), Eq.~\ref{pol}) at low gradients, while the opposite trend occurs in the solid regime. Fluid clusters exhibit high polarization, whereas solid clusters are dominated by rotation. 
  Dotted black lines mark entropy transitions, as in (d).
}
\label{FIG.main3}
\end{figure*}
\section{Results}
\subsection{ Mixing Entropy Indicates the Presence of Three Distinct Cluster Regimes}

In our simulations, we observed that clusters exhibited states with varying degrees of fluidity and particle rearrangements in the presence of a chemical gradient. To quantify the level of fluidity, we calculated the mixing entropy of the cluster. Particles were initially assigned three different colors, as shown in Fig.~\ref{FIG.main3}(a), and Supplementary Video M3, to track their spatial distribution and quantify mixing over time. We discretized the cluster into spatial bins (see Fig.~\ref{FIG.main3}(a)), where each bin represents a localized region containing a subset of particles. 
We then calculated the {\it weighted mixing entropy \( G \)} by considering particle densities in these discrete spatial bins using
\begin{equation}
  G = - \frac{\sum_{j=1}^{n_B} \rho(j) \sum_{i=1}^{3} p_i(j) \log(p_i(j))}{n_B}
\label{G}
\end{equation}
where, \( \rho(j) \) is density of particles in bin \( j \), \( p_i(j) \) is the probability of finding a particle of color \( i \) in bin \( j \) (\( i = 1, 2, 3 \)), \( n_B \) is the total number of bins used to discretize the cluster. The quantity \( G \) is the weighted Shannon entropy, representing the global mixing state of the cluster. 

The evolution of the mixing entropy over time for (\(N=37\)) is analyzed, comparing the fluid (\(g_r =0.0005\)) and solid (\(g_r =0.2\)) cases (see Fig.~\ref{FIG.main3}(b)). Our calculations, for a three color scheme, show that for a maximally mixed state, the mixing entropy is \(\ln{3}\), whereas for an ideal non-mixing case , the mixing entropy is expected to be zero, as particle positions remain fixed (see Appendix~\ref{SI.3}). However, due to finite system size effects and bin discretization constraints, we observe a systematic offset (\(\delta \sim 0.5\)) in the mixing entropy. While these effects diminish for larger systems, in our finite-sized simulations, they result in a nonzero baseline for the mixing entropy in the solid phase.
To account for this systematic offset, we use \(\delta\) as the baseline, leading to a corrected expectation where the mixing entropy for the completely mixed case becomes \(\ln(3) + \delta\) (see Fig.~\ref{FIG.main3}(b)). 
These baselines ensure a more accurate comparison of the mixing behavior across various gradients and parameters.


The mixing entropy as a function of the gradient is shown in Fig.~\ref{FIG.main3}(c), with vertical dashed lines marking the critical transition points defined as 10\% of the maximum saturated entropy and 10\% of the minimum entropy (see Fig.~\ref{FIG.main3}(d)).
We found that for a low gradient, the cluster is in a fluid regime in Fig.~\ref{FIG.main3}(c), characterized by high particle mobility and uniform mixing. In contrast, for high gradients, the cluster is in a non-mixing, solid-like regime, where particles exhibit minimal movement and maintain rigid configurations. This fluid and solid type behaviors are visually demonstrated in Fig.~\ref{FIG.main3}(e) and Fig.~\ref{FIG.main3}(g), and further captured in Supplementary Videos M3 and M5 respectively, which depict the time evolution of the color configuration within the clusters. The figures also denote the velocity vectors of each particle, further emphasizing the differences in dynamical behavior between these regimes.

In between the two vertical lines of Fig.~\ref{FIG.main3}(c), defined by the mixing entropy (Fig.~\ref{FIG.main3}(d)), we find a regime where the clusters deform and break up (Fig.~\ref{FIG.main3}(f) ; Supplementary Video M4.
We can therefore categorize the system into three distinct regimes: the fluid state characterized by complete mixing, the solid state with no mixing, and the intermediate breaking regime, which exhibits partial mixing and cluster break up. These phases and the breaking instability will be the focus of our study, but we first describe the surprising "solid" phase which appears in our model for strong gradients.

\subsection{Comparison of Fluid and Solid Type Phases.} 
\begin{figure}[htbp]
  \centering
\includegraphics[width=0.98\linewidth]{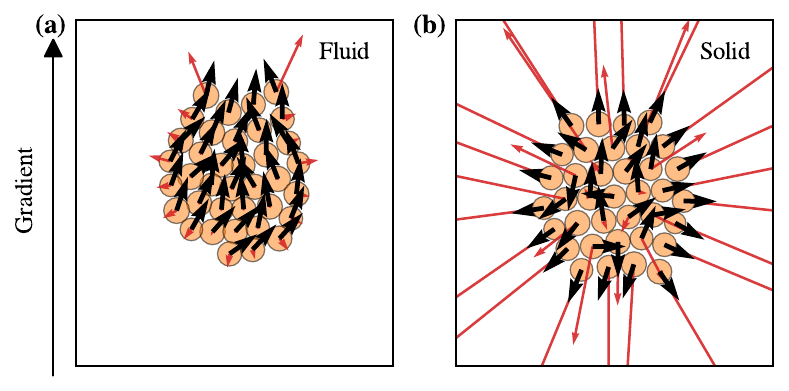}
\caption{\textbf{\bf Comparison of Fluid and Solid Type Phases.}
  (a) In the fluid phase, active propulsion unit vectors (black arrows, Eq. \ref{n}) exhibit weak alignment with the local-gradient vector (red arrows, Eq. \ref{g}) for low gradient slope. This results in a less-structured motion, where other interactions dominate.
(b) In the solid phase, active propulsion vectors strongly align with the local-gradient vectors for high gradient slope, leading to an outward-ordered arrangement that suppresses particle rearrangements and characterizes a solid-like state.
The simulations are performed with fixed parameters: $N=37$, and gradient values $g_r= 0.0005$ for the fluid phase and $g_r=0.2$ for the solid phase.}

\label{FIG.main4}
\end{figure}

The solid regime occurs under strong gradients where the cellular active propulsion forces act symmetrically (see Fig.~\ref{FIG.main4}(b), and Supplementary Video M8), creating a balanced force field that locks the cluster into a fixed, compact configuration. This transition is characterized by strong radial alignment of particle active propulsion vectors with the chemical gradient-induced vectors (see Fig.~\ref{FIG.main4}(b)), leading to an ordered outward arrangement and a solid-like state. Internal rearrangements are minimal and suppress the translational motion of the cluster (see Fig.~\ref{FIG.main3}(j)). 
In contrast, in the lower gradient regimes, weaker velocity alignment with the gradient vector (see Fig.~\ref{FIG.main4}(a), and Supplementary Video M6) results in a more fluid phase, where the motion of the particles remains less structured and is dominated by other interactions.
By considering response regimes where the system's behavior is determined by the dominance of either the self-propulsion and velocity alignment \( \hat{v}(t - dt) + \alpha \hat{V} \) or the applied gradient \( \beta \vec{g}_i \chi(y_i) \), we can identify a threshold gradient for rigidification, $g_{bs} \sim 0.015$, above which we expect solid clusters (see section.~\ref{SI.4.1} in Appendix~\ref{SI.4} for details).
That is, below the critical gradient value (\( g_r < 0.015 \)), the motion remains driven by self-propulsion and velocity alignment giving fluid clusters, while above this threshold (\( g_r > 0.015 \)), gradient-induced forces become dominant, stabilizing the rigid structure and suppressing motility.
The transition to a rigid structure reflects a loss of entropy, signaling a shift to an ordered state in which migratory capabilities are significantly diminished. This regime highlights how external forces above a critical amplitude can impose stability and rigidity, ultimately halting the dynamic behavior observed in the fluid phase.
Interestingly, while our simulations predict a solid phase at high gradient values, characterized by reduced cell rearrangement and increased rigidity, this regime was not observed under the experimental conditions tested. This discrepancy may arise from limitations in the experimental gradient range sampled or adaptive cellular behaviors, such as chemorepulsion at high gradients, that are not captured by our minimal model. As such, we interpret the emergence of the solid-like phase as a theoretical possibility within the model framework, which warrants further experimental investigation to confirm its physiological relevance.

\begin{figure*}[htbp]
\centering
\includegraphics[width=0.98\linewidth]{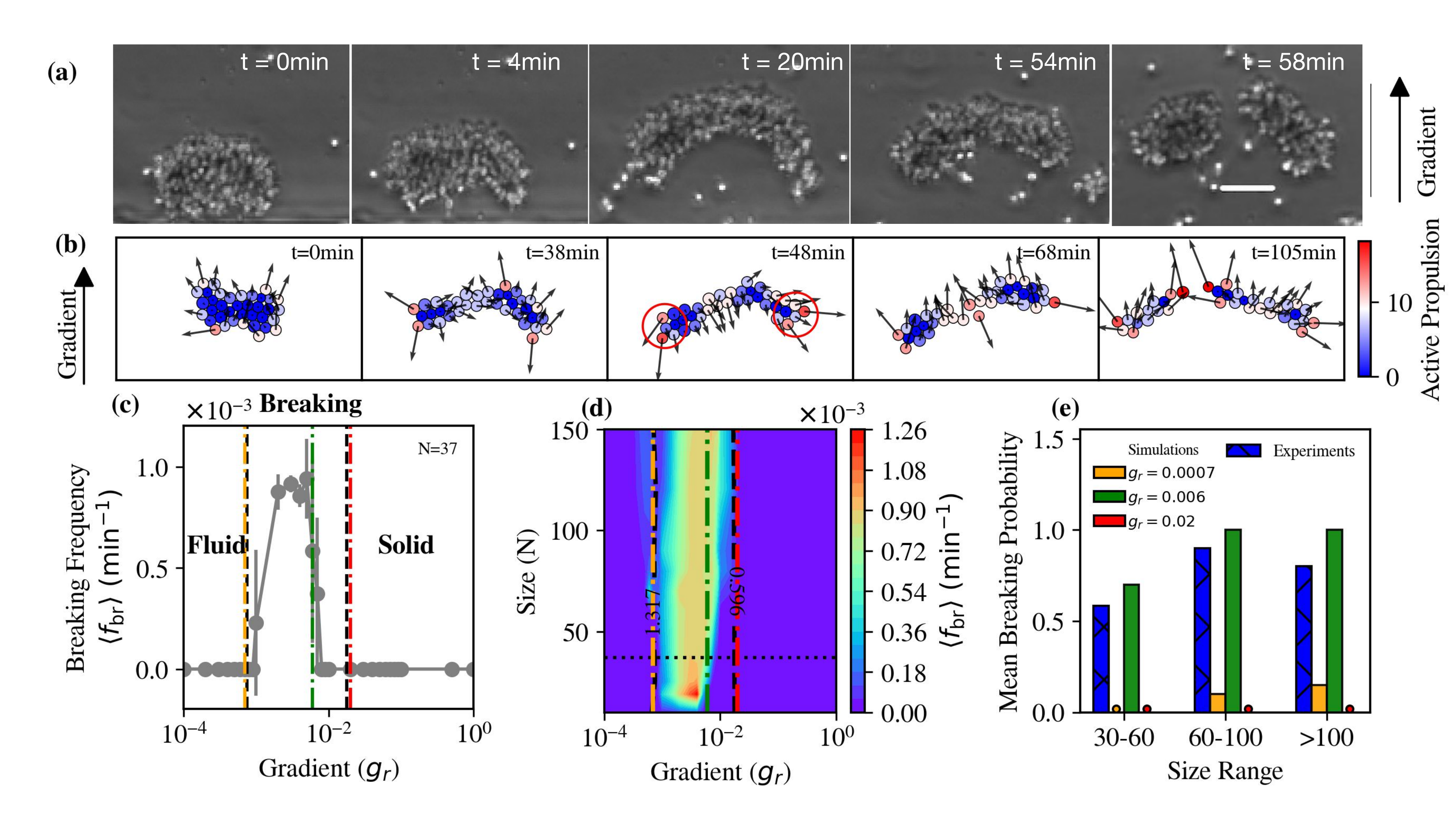}
\caption{\textbf{\bf Dynamics and Stability of Cluster Breakup: Experimental and Simulation Insights.}
(a) Time-lapse images of a cluster splitting into two, showing its transition from a symmetric to an elongated shape before fragmentation. A white line indicating a scale of 0.1 mm is included.
(b) Simulation snapshots of cluster breakup dynamics. Surface colors indicate active speed (red: high, blue: low), with arrows showing particle velocity. The sequence depicts a symmetric cluster elongating before fragmenting due to uneven velocity distribution, which amplifies local instabilities and disrupts cohesion. The external gradient is applied in upward the direction.
(c) Cluster breaking frequency rises with gradient in the breaking regime as force differences between the front and back accelerate fragmentation. As the gradient increases towards the solid regime, stabilization delays breakups. Vertical dashed lines mark critical transitions of entropy for N=37 (black) and indicate gradients $g_r= 0.0007$, $g_r= 0.006$, and $g_r= 0.02$ (red).
(d) Heatmap of average breaking frequency, showing its dependence on cluster size and gradient. The heatmap identifies the breaking regime, which aligns closely with the intermediate entropy regime which is indicated by the horizontal dashed line (see Fig.~\ref{FIG.main3}(d)). Vertical red dashed lines indicate gradients $g_r= 0.0007$, $g_r= 0.006$, and $g_r= 0.02$.
(e) Experimentally observed probability of cluster breaking as a function of cluster size, comparing small, medium, and large clusters, showing an increase of breakup probability with size (crossed histograms). Simulations for gradients $g_r= 0.0007$, $g_r= 0.006$, and $g_r= 0.02$ display histogram plots of average breakup probability by cluster size. Instances where the probability is zero are highlighted with small circles. The data demonstrate that clusters align closely with the breaking trend in the breaking regime. 
 }
\label{FIG.main5}
\end{figure*}

\subsection{ Forward Migration Analysis Reveals Optimal Range of Chemoattractant Gradient}
To quantify how effectively a cluster aligns its movement with an external chemical gradient, we calculated the Forward Migration Index (FMI), which
is defined as the ratio of the distance traveled by the cluster in the direction of the gradient to the total distance traveled. 
For gradients in the breaking regime, the FMI was calculated for clusters before they broke up. The FMI showed an initial linear increase with the gradient, followed by high fluctuations and a decline at very high gradient values (Fig.~\ref{FIG.main3}(h)). The initial linear increase in FMI suggests an optimal gradient range where clusters respond linearly to the externally applied gradient. This range is crucial because it represents a regime where clusters can most efficiently sense and respond to chemical signals. Such behavior is indicative of a balance between the driving gradient forces and the intrinsic cluster dynamics, enabling coordinated movement. Beyond this range, the observed fluctuations and decline in FMI may reflect the inability of clusters to maintain cohesion or directional migration under excessively strong gradients. 

Larger aggregates tend to have a higher ability to sense and respond to chemical signals more effectively \cite{malet-engraCollectiveCellMotility2015, copenhagen2018frustration}. This higher migratory efficiency can be attributed to the larger total force due to the larger chemotactic force difference across the diameter of the larger clusters \cite{copenhagen2018frustration}, which enable them to maintain directional movement despite environmental fluctuations. To quantify this trend, we examined how cluster size influences response dynamics under varying gradients in our model. A heatmap of FMI as a function of size and gradient (see Fig.~\ref{FIG.main3}(i)), reveals that larger clusters exhibit more effective migration than smaller ones at low gradients, as one might expect. 
Indeed, the trend of the size dependence of FMI in the breaking regime for $g_r = 0.006$ (see Fig.~\ref{FIG.SI2}) aligns closely with the observed size dependence of FMI in the experiments (see Supplementary Figure~SI2($D^{ii}$) of \cite{malet-engraCollectiveCellMotility2015}).
 Interestingly, in our model, this trend continues at the high gradients of the solid phase, despite the overall decrease in the average speed of the clusters (Fig.~\ref{FIG.main3}(j)). 

\subsection{Fluid Clusters Maintain High Polarization While Solid Clusters Tend to Rotate}

The internal dynamics of clusters strongly affect their overall migration behavior \cite{malet-engraCollectiveCellMotility2015, copenhagen2018frustration}. To investigate these aspects, we analyzed the angular momentum, and polarization of clusters. 
 Angular momentum quantifies a cluster's rotational motion based on individual particle movements around its center of mass. We compute the mean absolute angular momentum per particle as:
\begin{equation}
\mathcal{A}_t = \left|\frac{1}{T} \sum_{t=1}^{T} \mathcal{A} \right| 
\label{a_max}
\end{equation}
 where,
\begin{equation}
\mathcal{A} = \frac{1}{N} \sum_{i=1}^{N} \vec{r}_{i,cm} \times \vec{v}_{i,cm}.
\end{equation}
Here the position and velocity of a particle relative to the center of mass are given by \(\vec{r}_{i,cm} = \vec{r}_i - \vec{r}_{\text{cm}}\) and \(\vec{v}_{i,cm} = \vec{v}_i - \vec{v}_{\text{cm}}\), with \(\vec{r}_i\) and \(\vec{v}_i\) representing the particle’s position and velocity, and \(\vec{r}_{\text{cm}}\), \(\vec{v}_{\text{cm}}\) denoting the center of mass position and velocity of the cluster. This average provides a measure of the average rotational motion of the particles over time $T$. 
In addition to \(\mathcal{A}_t\), polarization measures velocity alignment within the cluster, capturing the degree of collective translational order \cite{copenhagen2018frustration}:
\begin{equation}
\mathcal{O} = \frac{1}{N} \left| \sum_{i=1}^{N} \frac{\vec{v}_i}{|\vec{v}_i|} \right|
\label{pol}
\end{equation}
where \( {\vec{v}_i}/{|\vec{v}_i|} \) is the unit velocity vector of particle \( i \). A high polarization value (\(\mathcal{O} \approx 1\)) indicates strong velocity alignment, whereas a low value (\(\mathcal{O} \approx 0\)) suggests disordered motion.

The dynamical properties of clusters evolve with gradient strength, reflecting shifts in polarization and rotational dynamics. At low gradients, fluid clusters exhibit high translational motion (see Fig.~\ref{FIG.main3}(j)) and polarization, facilitating mixing, while rotational dynamics remain low. As the gradient increases, polarization declines, destabilizing fluid clusters (Fig.~\ref{FIG.main3}(k)). In the high-gradient regime, clusters become rigid and dominated by angular rotation, with low polarization (as seen in Fig.~\ref{FIG.main4}(b)), and limited particle exchange. The intermediate regime shows a gradual transition, where polarization decreases and rotational displacement increases, highlighting the link between gradient strength and cluster behavior. 


\subsection{Cluster breaking regime}

\begin{figure}
  \centering
  \includegraphics[width=1\linewidth]{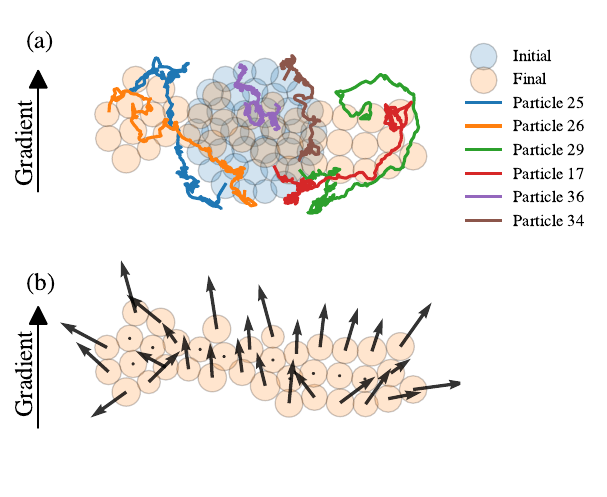}
  \caption{{\bf (a) Trajectories of selected particles in a migrating cluster.} The initial and final positions of all particles are represented as translucent blue and orange circles in COM frame, respectively. The paths of four tracked particles from the rear end of the cluster and two front front of the cluster are highlighted in distinct colors, showing their movement from the initial to the final frame over the simulation duration. This illustrates the process of cluster elongation perpendicular to the gradient direction, as particles are pulled by the gradient to move around those in front of them, since cells at the two sharp corners are mostly pulling the cluster sideways. {\bf (b) Final cluster configuration with propulsion directions.} Particle positions at the final frame are shown, with arrows indicating their active propulsion directions, illustrating the collective migration and lateral elongation of the cluster.}
  \label{FIG.main6}
\end{figure}


A striking observation in both experiments and simulations is the elongation and fragmentation of clusters under intermediate gradient conditions. Experimental time-lapse imaging shows stable clusters at low gradients (0-50 ng/ml/mm), (Supplementary Video M9) whereas for intermediate gradients (0-100 ng/ml/mm), we see clusters breaking into multiple smaller clusters over time (Fig.~\ref{FIG.main5}(a) and Supplementary Video M10). Similar dynamics are observed in simulations with stable clusters at low gradients (Supplementary Video M6) and instability at intermediate gradients (Fig.~\ref{FIG.main5}(b); Supplementary Video M11), where cluster breakup closely mirrors the experimental observations. The process of cluster elongation and eventual breakup is driven by cellular active propulsion acting on the rim particles of the cluster. 
Particles at the front of the cluster exhibit high persistence along the gradient, maintaining cohesive motion, while rear particles experience lower persistence. Particles are primarily pulled by the gradient when they "peek" to the left or right of those ahead. The gradient's pull increases, giving them more space to move directly upwards (Fig.~\ref{FIG.main6}(a), Supplementary Video M12). This movement flattens the leading edge and leads to the breaking of the cluster’s symmetry. As symmetry breaks, the active velocity of rim particles that reside at the high-curvature right and left edges of the cluster increase further, stretching the cluster sideways (Fig.~\ref{FIG.main5}(b)). This feedback loop elongates the cluster, causing instability and fragmentation.

To gain a quantitative understanding of the breaking transition, we analyzed a variant of our model system where particle positions were fixed but all other quantities evolved normally (see Appendix~\ref{SI.7}). We measured the correlation between active propulsion vectors as a function of distance under varying gradients. As the cluster transitions from a low gradient (fluid state for motile clusters) to an intermediate gradient (breaking regime for motile clusters), we observed that the correlation length (measured by the distance at which the correlation vanishes in the correlation plot (see Fig.~\ref{FIG.SI3}(a))) shifts from approximately 5.1~$r/\zeta$ in the fluid regime to around 2.7~$r/\zeta$ in the breaking regime. Here, $\zeta = \langle \sigma_{ij} \rangle$, is the averaged particle diameter. This shift indicates a decrease in the correlation of active propulsion vectors from the fluid to the breaking regime. Additionally, an averaged analysis of the correlation length as a function of gradients showed that the correlation remains consistently lower in the breaking regime compared to the fluid regime (see Fig.~\ref{FIG.SI3}(b) and Appendix~\ref{SI.7}).


To quantify the fragmentation dynamics, we measured the frequency of breaking of clusters in different gradient regimes (Fig.~\ref{FIG.main5}(c)). The breaking frequency \( f_{\text{br}} \), defined as the inverse of the breaking time, quantifies the rate at which  clusters fragment into smaller clusters. Here clusters are defined as groups containing more than three particles and a value of \( f_{\text{br}} =0\), indicates no fragmentation.
We find that the breaking frequency initially increases with gradient strength, as larger chemotactic forces induce faster cluster stretching, elongation, and breaking. Using similar arguments to the ones used to compute the threshold gradient for rigidification, $g_{bs}$, we identified a critical gradient (\( g_{fb} \)), where clusters transition from a flocking-dominated to a breaking-dominated regime (see Appendix~\ref{SI.4}). In this regime, the gradient becomes strong enough to overcome the self-persistence of individual particles, leading to fragmentation. However, the global alignment remains sufficiently high to allow the formation of smaller, localized domains. Thus, the breaking regime should lie within the range \( g_{fb} < g_r < g_{bs} \), where both self-propulsion and alignment effects compete with the external gradient influence. However, as the gradient approaches the solid regime transition (\( g_{bs} \)), stabilization effects begin to counteract fragmentation, delaying breakup events and reducing the overall breaking frequency.
This suggests that while intermediate gradients promote transient fragmentation, stronger gradients beyond \( g_{bs} \) lead to more rigid structures dominated by rotational motion, ultimately suppressing cluster turnover.

Cluster stability is also influenced by size, as observed in both simulations and experiments. A heatmap of breaking frequency (Fig.~\ref{FIG.main5}(d)) reveals that larger clusters are more prone to instability compared to smaller ones, primarily due to amplified chemotactic force differences acting across their structure. 
This size dependence of fragmentation is further confirmed by analyzing the probability of cluster breakup as a function of size. Using simulations we calculate the average breakup probabilities
for small, medium, and large clusters, for gradients $g_r= 0.0007$, $g_r= 0.006$, and $g_r= 0.02$ (Fig.~\ref{FIG.main5}(e)). The simulations show a clear trend: larger clusters exhibit a significantly higher probability of fragmentation, with good agreement between our simulation data in the breaking regime ($g_r= 0.006$) and experimental data obtained primarily for intermediate gradients (0-100 ng/ml/mm). 
These findings suggest that as gradients strengthen, larger clusters become increasingly vulnerable to breakup due to dynamic instability, particularly within specific gradient ranges.

\begin{figure*}[htbp]
  \centering
  \includegraphics[width=0.99\linewidth]{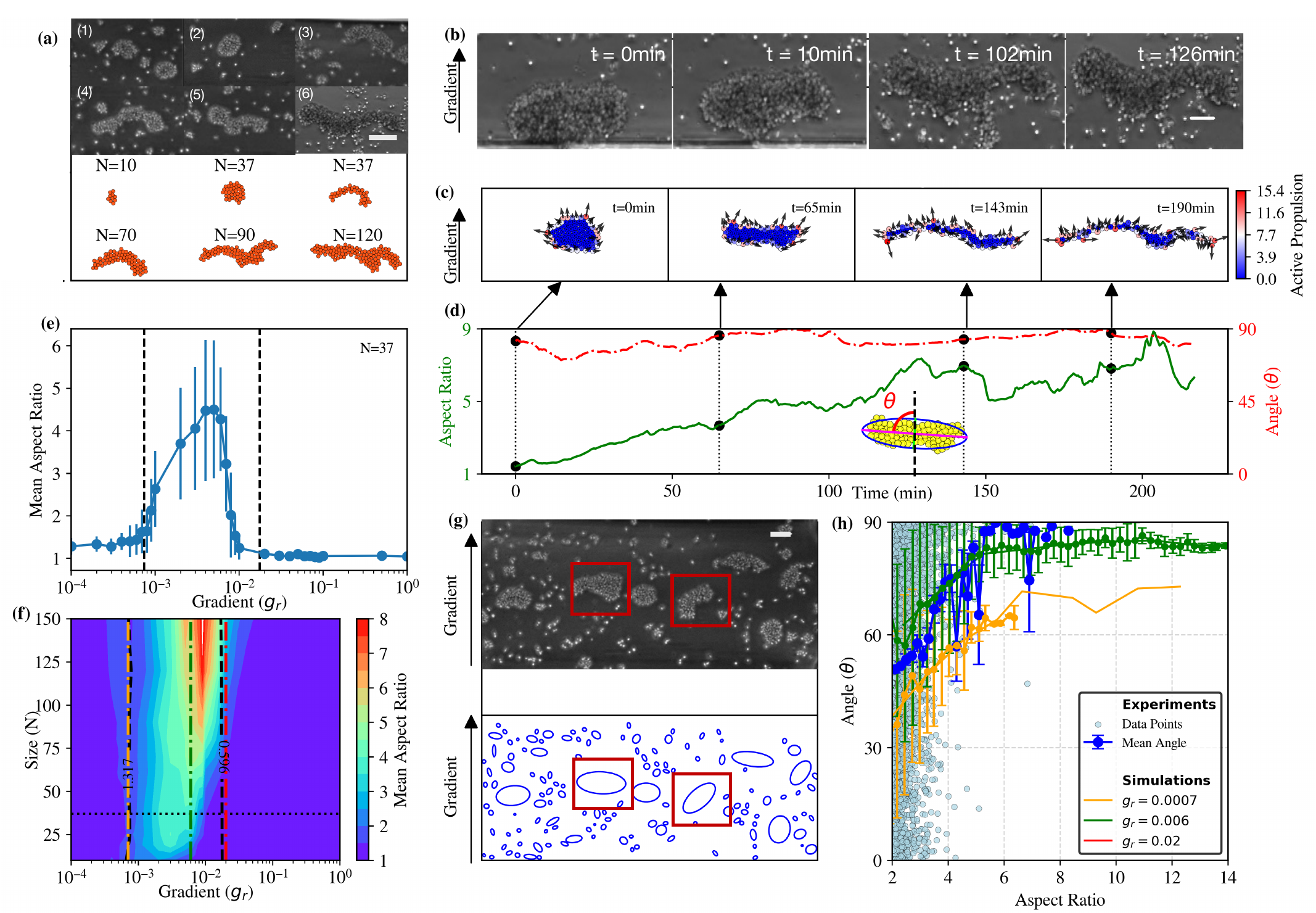}
  \caption{{\bf Orientation and Shape Dynamics in Gradient-Driven Cluster Breakup.}
  (a) Panel 1: Experimental images of migrating clusters showing an increase in size, as illustrated in figure from (1) to (6). The images capture both steady-state (stable) and transient (unstable or deforming) shapes as the clusters respond to external gradients (direction of gradient to +y axis). Scale bar = 0.1 mm.
  Panel 2: Simulated cluster of different sizes with shapes similar to the experimental observations (direction of gradient to +y axis).\
  (b) Experimental image of a large cluster elongating and forming wing-like structures prior to breaking. A white line indicating a scale of 0.1 mm is included.\
  (c) Simulation snapshots illustrating the formation of wing-like structures in a large cluster prior to breaking.\
  (d) Simulation results showing the time evolution of cluster orientation angle and aspect ratio prior to breakup. 
  To illustrate structural changes, the major (magenta) and minor (lime) axes of the cluster (yellow) are outlined, and an ellipse (blue) is fitted around it. The angle between the major axis and the positive y-axis (red) is emphasized, aligning with the calculated values presented in the plot. The presented dynamics correspond to a gradient of $g_r= 0.005$ and a cluster size of $N=90$.
  (e) Mean aspect ratio as a function of gradient, highlighting significant shape anisotropy and variability during the cluster-breaking regime, with stability observed in high-gradient regions. Dotted black lines mark entropy-based transitions (see Fig.~\ref{FIG.main3}(d)).\
  (f) Heatmap of aspect ratio as a function of size and gradient, illustrating shape anisotropy across different cluster states within the breaking regime. \(N = 37\) is marked by a horizontal dashed line, with entropy contours (see Fig.~\ref{FIG.main3}(d)).
  (g) Experimental snapshots of cluster systems, with selected clusters highlighted in red boxes. A white line indicating a scale of 0.1 mm is included. The bottom panels show fitted ellipses over clusters, emphasizing structural features and shape anisotropy.\
  (h) Scatter plot of experimental cluster orientation angle ($\theta$) versus aspect ratio, revealing that elongated clusters tend to align orthogonally to the direction of motion. The blue line represents the average orientation angle for each aspect ratio. Simulation results for gradients, $g_r= 0.0007$, $g_r= 0.006$, and $g_r= 0.02$ show close agreement with experimental data in the breaking cluster regime ($g_r= 0.006$).\
}
 \label{FIG.main7}
\end{figure*}

\subsection{Cluster Elongation Orthogonal to the Direction of Motion}
As shown above, our simulations predict that cluster breaking is driven and preceded by the tendency of clusters to elongate in the direction orthogonal to the chemoattractant gradient. To test this prediction and investigate the effect of cluster size on migration and stability, we analyzed both experimental and simulated shapes of various sizes with various parameters (Fig.~\ref{FIG.main7}(a)). Simulated cluster shapes strongly resemble experimental observations, validating the model and providing a reliable foundation for further analysis. This resemblance enables us to explore how structural variations influence cluster dynamics, particularly in the breaking regime.
The time evolution of large cluster is illustrated through experimental images (Fig.~\ref{FIG.main7}(b), Supplementary Video M13), complemented by snapshots from simulations (Fig.~\ref{FIG.main7}(c)\, Supplementary Video M14). Comparative analyses revealed that both experimental and simulated clusters undergo significant elongation prior to breakup, with their major axes aligning nearly orthogonally to the direction of migration (see Fig.~\ref{FIG.main5}(a,b). This alignment suggests a structural transition associated with cluster destabilization, where increasing asymmetry contributes to the onset of fragmentation.

To further quantify this relationship, we examined the aspect ratio and orientation angle with respect to the chemoattractant gradient direction over the same time series as depicted in Fig.~\ref{FIG.main7}(c), presenting the results in Fig.~\ref{FIG.main7}(d). The aspect ratio quantifies the elongation of a cluster by comparing its dimensions along the principal axes. It is defined as the ratio of the square roots of the eigenvalues of the gyration tensor,
\(
 \sqrt{{\lambda_1}/{\lambda_2}}, \quad \text{where } \lambda_1 \geq \lambda_2.
\)
An aspect ratio close to 1 indicates a symmetric, circular shape, while higher values suggest an elongated or anisotropic cluster shape.
The cluster orientational angle (\( \theta \)) is defined as the angle between the major axis of the cluster and the direction of the gradient (positive \( y \)-axis). This angle is calculated from the eigenvectors of the gyration tensor. If \( \mathbf{v}_1 = (v_{1x}, v_{1y}) \) is the eigenvector corresponding to the largest eigenvalue \( \lambda_1 \), then the angle is given by:
\(
\theta = \arctan\left(\frac{v_{1x}}{v_{1y}}\right)
\)
(see inset of Fig.~\ref{FIG.main7}(d)).
Shape analysis reveals critical structural insights into clusters approaching the breaking regime. We focus on the aspect ratio as a key metric to quantify these structural changes. Our analysis shows that the aspect ratio of clusters increases significantly in the breaking gradient region (Fig.~\ref{FIG.main7}(e)). In this regime, clusters become highly anisotropic, stretching into elongated shapes with substantial fluctuations. These findings establish a clear link between shape asymmetry and instability, with clusters in low-gradient and high-gradient regions exhibiting minimal shape variations.

To further highlight these findings, we present a heatmap of aspect ratio as a function of size and gradient (Fig.~\ref{FIG.main7}(f)). This heatmap vividly captures the anisotropic nature of clusters across different states within the breaking regime. 
The heatmap clearly shows high aspect ratio peaks corresponding to larger cluster sizes. Additionally, for clusters of greater size, anisotropy begins to emerge at lower gradient values and persists across higher gradients, emphasizing the early onset of structural changes in larger clusters and their prolonged influence under varying conditions.

This orthogonal alignment during fragmentation arises from the interplay between gradient-induced persistence and shape fluctuations. Experimental data reveals a strong correlation between aspect ratio and alignment. To quantify size parameters, such as the major and minor axes and the orientation of clusters in experimental images (e.g., Fig.~\ref{FIG.main7}(g)), we used Fiji’s \textit{Measure} feature. The extracted data from Panel 1 are plotted in Panel 2, and this quantified data is then used to analyze experimental trends in Fig.~\ref{FIG.main7}(h). 
Scatter plots of experimental cluster orientation angle (\(\theta\)) versus aspect ratio indicate that elongated clusters tend to align orthogonally to the direction of motion (Fig.~\ref{FIG.main7}(h)). The guided red line represents the average orientation angle for each aspect ratio, illustrating a clear trend of orthogonal alignment with increasing elongation. Simulations conducted for gradients \( g_r= 0.0007 \), \( g_r= 0.006 \), and \( g_r= 0.01 \) show strong agreement with experimental data in the breaking cluster regime (\( g_r= 0.006 \)). Fig.~\ref{FIG.SI4} highlights how cluster orientation varies with gradient strength, particularly for elongated clusters (aspect ratio \( > 2 \)), emphasizing the tight coupling between orientation dynamics and the prevalence of elongated clusters in the breaking regime.

To understand whether there could be any functional advantages to clusters residing in the breaking regime, we examine the cluster's response to the gradient, characterized by its velocity component in the gradient direction, i.e., velocity in the Y direction. 
The resulting velocity in the Y direction reaches its highest levels in the breaking regime (see Fig.~\ref{FIG.SI5}(a,b)). This enhanced response suggests that clusters in the breaking regime may benefit from greater adaptability and agility in dynamic {\it in vivo} environments.

\section{Discussion}

Our study provides insights into the chemotactic efficiency, behavioral transitions, and breakup dynamics of migrating cellular clusters during collective chemotaxis. Using experimental and simulation data, we characterized the influence of external chemical gradients on cluster behavior.


Our results show a non-monotonic entropy trend with gradient intensity, revealing distinct behavioral regimes. In low-gradient conditions, clusters exhibit high mixing entropy and fluid-like behavior (Fig.~\ref{FIG.main3}(e)). At intermediate gradients, entropy decreases as clusters elongate and partially mix, marking a transition phase (Fig.~\ref{FIG.main3}(f)). Under high gradients, clusters become rigid, with rotational dominance and minimal mixing (Fig.~\ref{FIG.main3}(g)). 
While our computational model predicts a transition to a rigid, solid-like cluster phase at very high chemoattractant gradients, such a phase was not observed in our experiments even at very high gradients. One potential reason is that the lymphocytes in our system undergo chemorepulsion (or fugetaxis) under extreme gradient conditions, particularly via CCR7 signaling pathways \cite{copenhagen2018frustration, malet-engraCollectiveCellMotility2015}. This phenomenon is quite general and has been observed in various immune cells, where excessively high concentrations of chemokines like CCL19 or CXCL12 induce a reversal in migration direction \cite{poznansky2000}. Such chemorepulsive effects will alter the dynamics of the cells thereby obscuring the predicted rigid phase.


Another possibility is that the rigid phase may indeed exist, but only within a very narrow window of chemoattractant gradient strengths, before the onset of chemorepulsion, that was not sampled experimentally . Given that the transition from fluid-like to solid-like collective states can be abrupt, even small changes in gradient magnitude could shift the system across phase boundaries. It would be interesting to do systematic gradient scans in future experiments to capture potential rigid phases, potentially with systems that do not show chemorepulsion. It is also interesting to contrast this sort of rigidity transition driven by an external signal to jamming/unjamming transitions in epithelial sheets that are related to changes in the cell shape, density and adhesion. \cite{park2015unjamming,Ilina2020,nir2015physics}


Our work highlights an intriguing feedback-driven breakup instability in chemotactic lymphocyte clusters exposed to intermediate chemoattractant gradients (Fig.~\ref{FIG.main5}(a, b)). We showed that, beyond a threshold gradient, clusters elongate perpendicular to the gradient and eventually fragment into smaller clusters. This instability is driven by a feedback mechanism where the front-edge cells move persistently up the gradient while cells toward the sides slip forward and flatten the leading edge, breaking the cluster’s symmetry. The high curvature sides then experience increased persistent motion outwards, stretching the cluster laterally until it breaks (Fig.~\ref{FIG.main6}). 
Consistently, we found that the probability of breakup rises with gradient strength and with cluster size (Fig.~\ref{FIG.main5}(d), (e)), indicating a size-dependent instability threshold that disrupts cluster cohesion and limits cluster size.

From a physics perspective, this behavior can be viewed in the context of the process of equilibrium phase separation due to attractive adhesion interactions between particles which is disrupted by activity, that has been studied in different synthetic systems ranging from colloids to active nematics \cite{aparna_2013reentrant, tayar2023}. However, in our system, the activity is not uniform and its influence is coupled to the boundary shape.  One can also consider extending this concept to the context of motility-induced phase separation (MIPS) in active matter systems, where, in the absence of external cues, self-propelled particles with purely repulsive interactions can spontaneously phase-separate into dense clusters and dilute gas-like regions \cite{cates2015motility,sanoria_influence_2021,sanoria_percolation_2022}. Such coarsening would typically produce a single large cluster at steady state. However, in our chemotactic system the external gradient provides a directed feedback that produces an instability that causes fragmentation and counteracts indefinite coarsening which may be observable in appropriately designed active colloidal systems \cite{dauchot2019interrupted}. 


The elongated, wing-like shapes that clusters adopt before breaking are seen in both our simulations and experiments (Fig.~\ref{FIG.main7}(b, c)) and are reminiscent of fingering instabilities observed in other active media. In spreading cell monolayers, for instance, an advancing tissue front often develops finger-like protrusions due to an interfacial instability driven by active forces \cite{alert2019active}. It is interesting to note that, contrary to typical fingering instabilities where fingers grow in the gradient direction, the stretching of clusters and their eventual break-up occurs in a direction perpendicular to the gradient as shown by our aspect ratio and orientation measurements in both simulations and experiments Fig.~\ref{FIG.main7}(d-h)).


 
 Biologically, the fragmentation of large clusters under strong chemotactic drive may represent a functional mechanism to constrain cluster size and enhance adaptability. 
 There is growing evidence that collective dissemination — small groups of cells traveling together — is a highly effective route for metastasis \cite{cheung2016}, suggesting that intermediate-sized clusters balance efficiency of spread with survivability. Indeed, theoretical and experimental studies indicate that clusters retain certain advantages (such as cell–cell cooperation and survival signals) but avoid the drawbacks of excessive size \cite{cheung2016}. In this light, the feedback-driven breakup instability we observe could be biologically advantageous: by breaking a large cluster into smaller ones, the cells might maximize migratory efficiency and dispersion in complex environments. 
Interestingly, comparing the model to experiments, we find that cell clusters appear to reside in the break-up regime for gradients that are comparable with physiologically relevant ones \cite{ malet-engraCollectiveCellMotility2015}, We explicitly showed that functional benefits in this regime includes enhanced speed and a stronger response to the gradient, as suggested by the increased velocity in the Y direction observed in this regime. We also showed that, within this regime, migration in response to the gradient is fastest for cluster sizes that stay cohesive, i.e., below a critical size (see Fig.~\ref{FIG.SI5}). This suggests that cells could even tune their chemotactic response strength to stay in this functionally beneficially regime.
Our findings thus point to a potential mechanism by which chemotactic cues and the cellular responses to them inherently regulate cluster size in vivo – preventing clusters from becoming too large to traverse the environment, and possibly optimizing the trade-off between cohesive group migration and the need to disperse. 



In summary, by identifying behavioral regimes, stability thresholds, and shape transformations of chemotactically driven multicellular clusters, we provide a framework for understanding transitions between fluid-like, breaking, and rigid phases. The feedback-driven cluster instability reported here connects to broader themes in active matter and biology and it suggests that nature may utilize such instabilities to maintain optimal cluster sizes for collective migration and function. Our work is relevant to collective cell migration, where chemical gradients impact cohesion and motility and provides deeper insights into cluster stability and fragmentation mechanisms, with relevance to bacterial colonies, synthetic colloids, and cellular aggregates as well as informing active material design for a wide range of applications.

\section*{MATERIALS AND METHODS}

The human chronic lymphocytic leukemia (CLL)-derived cell line JVM3 was obtained from the DSMZ (Deutsche Sammlung von Mikroorganismen und Zellkulturen) and routinely authenticated via B cell surface marker analysis, CCR7 expression, and mycoplasma testing. Recombinant human CCL19 chemokines were purchased from PeproTech, and Alexa Fluor 647–labeled CCL19 from Almac; both were aliquoted and stored according to the manufacturers’ instructions.
Cell motility in response to CCL19 gradients was assessed by video microscopy using collagen IV–coated 2D chemotaxis slides (ibidi), as previously described~\cite{malet-engraCollectiveCellMotility2015}. Briefly, $0.5 \times 10^6$ cells in 10~ml culture medium were loaded into the central channel of a Collagen-IV-treated 2D chemotaxis µ-slides from Ibidi and incubated at 37$^\circ$C for 30 minutes to allow adhesion. Chemokine gradients were established according to the manufacturer’s protocol and verified for linearity using a 10\% dextran–fluorescein isothiocyanate solution.
Cell migration was recorded with wide-field microscopes equipped with environmental control chambers (temperature and CO$_2$), acquiring one frame every 2 minutes using MetaMorph software (Molecular Devices)~\cite{malet-engraCollectiveCellMotility2015} on either a Zeiss Axiovert 200 microscope (2.5x objective, coolSNAP HQ CCD camera, Photometrics) or a Nikon Eclipse TE2000-E microscope (4x/0.13 NA objective, Cascade II 512 CCD camera, Photometrics).
Cluster shape and orientation were analyzed from the same time-lapse videos using ImageJ's built-in \texttt{Measure} function. Each cluster was fitted to an ellipse, and the orientation angle of the major axis was extracted. Breaking events were identified by monitoring the size of individual clusters over time.


\appendix
\renewcommand{\thefigure}{SI\arabic{figure}}
\setcounter{figure}{0}
\section{Conversion of Simulation Time Steps to Experimental Time}
\label{SI.1}
To align simulation data with experimental observations, we compared average particle speeds and sizes. Using a midpoint experimental particle radius of $15\,\mu\text{m}$ and a simulation radius of 30 units and the experimental speed of $5\,\mu\text{m}/min$ and the average simulation speed of $5 \, \text{units}/ \text{time step}$, this yields a time conversion factor of 1 simulation time step = 30 seconds. Consequently, 4000 simulation time steps represent approximately 33.3 hours of experimental time.

\section{Neighbor Interaction Cutoff Definitions}
\label{SI.2}

We use a two-shell approach to define particle neighborhoods. The \textbf{first-nearest neighbor shell} includes particle pairs close enough for interparticle forces to act. The interaction range is based on the characteristic size of the pair, $\sigma_{ij}$, for particles $i$ and $j$ (see Eq.~\ref{LJ}), and is defined by:
\begin{equation}
d < \sigma_{ij} + \Delta_1 
\label{firstCutOff}
\end{equation}
where $d$ is the center-to-center distance between particles, and $\Delta_1$ provides a uniform buffer beyond contact. We set $\Delta_1 = R_1 - \zeta$, with $R_1 = 38$ and $\zeta = 30$ as the nominal particle diameter, yielding $\Delta_1 = 8$. For uniform particle size, this cutoff corresponds to approximately $1.25\zeta$ and effectively captures all particles subject to short-range interactions.

The \textbf{second-nearest neighbor shell} includes particles outside the first neighbor range but within a broader zone of influence. It is defined by:
\begin{equation}
\sigma_{ij} + \Delta_1 < d < \sigma_{ij} + \Delta_2 
\label{secondCutOff}
\end{equation}
where $\Delta_2 = R_2 - \zeta$ gives the extension of the second shell. With $R_2 = 100$ and $\zeta = 30$, we set $\Delta_2 = 70$. This two-shell method efficiently identifies neighbors and accommodates slight variations in particle size.

\section{ Mixing Entropy of a Three-Layer Particle System}
\label{SI.3}
We consider a system of particles arranged in a square lattice, divided into three colored distinct horizontal layers (e.g. blue, white, red) . Here we examined two special cases where these particles can either remain within their respective layers (unmixed) or completely mix across all three layers. We calculate the entropy for both cases.
\subsection{Entropy of the Completely Unmixed Case}
When the three color particle layers are completely unmixed, the particles remain confined to their specific layers. 
In this scenario, the probability $p_i$ for each particle to be in its designated layer is 1, and the probability of being in any other layer is 0.

Using the entropy formula:
\begin{equation}
S = -k_B \sum_{i} p_i \ln(p_i)
\end{equation}
Since $p_i = 1$ for each unmixed layer:
\begin{equation}
S = -k_B (1 \cdot \ln(1)) = 0
\end{equation}
This result indicates zero entropy because there is no uncertainty or disorder; particles are perfectly ordered and localized in their own layers without any mixing.

\subsection{Entropy of the Completely Mixed Case}
When the three color particle layers are \textbf{completely mixed}, the particles are spread out evenly across all three layers.
In this scenario, the probability $p_i$ for any particle to be in a particular layer is equal for all layers. Since there are three layers, each particle has a \textbf{$p_i = \frac{1}{3}$} chance of being in any given layer.
Using the entropy formula:
\begin{equation}
S = -k_B \sum_{i} p_i \ln(p_i)
\end{equation}
Since $p_i = \frac{1}{3}$ for each layer:
\begin{equation}
S = -k_B \left( \frac{1}{3} \ln\left(\frac{1}{3}\right) + \frac{1}{3} \ln\left(\frac{1}{3}\right) + \frac{1}{3} \ln\left(\frac{1}{3}\right) \right)
\end{equation}
\begin{equation}
S = -k_B \left( 3 \times \frac{1}{3} \ln\left(\frac{1}{3}\right) \right)
\end{equation}

\begin{equation}
S = -k_B \ln\left(\frac{1}{3}\right)
\end{equation}
Since $\ln(1/3) = -\ln(3)$, we get:
\begin{equation}
S = k_B \ln(3)
\end{equation}

This means that the \textbf{mixing entropy is maximized}, as there is the highest possible uncertainty and disorder—particles are no longer restricted to their original layers and are equally likely to be found anywhere.

\section{Threshold for Fluid, Breaking and Solid-like Regime}
\label{SI.4}

From Equations \ref{v} and \ref{n}, neglecting the terms \(\vec{LJ}\) and \(\vec{S}\), the velocity evolves as:
\[
\vec{v}_i(t - \Delta t) \approx p_i \hat{n}_i(t - \Delta t) + \eta,
\]
where \(\hat{n}_i\) is the unit vector describing the direction of motion, and \(\eta\) represents noise.

The direction vector \(\hat{n}_i\) is given by:
\[
\hat{n}_i = \frac{\vec{s}}{|\vec{s}|}, \quad \text{where} \quad \vec{s} = \hat{v}_i(t - \Delta t) + \alpha \hat{V} + \beta \vec{g}_i \chi(y_i),
\]
where \(\hat{v}_i(t - \Delta t)\) is the velocity of the particle at the previous time step reflecting persistence. Alignment is reflected by \(\alpha\), the strength of the alignment interaction, and \(\hat{V}\), the unit vector of the mean orientation of nearest-neighboring cells. Finally the response to the local concentration of chemoattractant at the location of the cell, \(\beta\chi(y_i)\), and the local gradient vector, $\vec{g}_i$, together reflect the influence of the chemoattractant.

We consider that the noise is small and the system is dominated by active propulsion, hence by \(\hat{n}_i\). The direction of \(\hat{n}_i\) is primarily determined by the largest term in \(\vec{s}\), given that \(\hat{n}_i\) is a unit vector. The magnitudes of the terms in \(\vec{s}\) are:

\begin{itemize}
  \item \textbf{Active propulsion term:} \( |\hat{v}_i(t - \Delta t)| = 1 \)
  \item \textbf{External velocity influence:} \( |\alpha \hat{V}| = \alpha = 6 \) (since \(\hat{V}\) is a unit vector)
  \item \textbf{Gradient influence:} \( |\beta \vec{g}_i \chi(y_i)| \approx \beta |\vec{g}_i| \chi(y_i) \), where \( \beta = 80 \), and \( |\vec{g}_i| \approx 1 \)
  \item \textbf{Response function:} \( \chi(y_i) = \frac{y_0 \cdot g_r}{y_0 \cdot g_r + 40} \), from Eq.\ref{cy_sat}, where \(c(y_i) \approx y_0 \cdot g_r\), and \(y_0 = 250\)
\end{itemize}

Based on these magnitudes, we now note that \( \hat{v}_i(t - \Delta t) < \alpha \hat{V}\) and that, depending on the value of $g_r$, the gradient term, \( \beta \vec{g}_i \chi(y_i) \), could lie below, between or above the two values of the persistence and alignment terms, potentially giving us three phases.

\subsection{High Gradient Regime $(g_r > g_{bs})$}
\label{SI.4.1}
As we lower the value of $g_r$ from very high values, a first critical point occurs when the alignment term \(  \alpha \hat{V}  \) and the gradient term balance \( \beta \vec{g}_i \chi(y_i) \), i.e., when:
\[
\alpha \hat{V} \approx \beta \vec{g}_i \chi(y_i).
\]

Using the values quoted above and solving for the critical gradient \( g_{bs} \), we obtain:
\[
80 \frac{250 \cdot g_{bs}}{250 \cdot g_{bs} + 40} = \alpha,
\]
\[
g_{bs} = \frac{40(\alpha)}{20000 - 250(\alpha)} = \frac{280}{18250} \approx 0.013.
\]

This indicates that for \( g_r < 0.013 \), the motion is dominated by the collective velocity alignment \( |\alpha \hat{V}| \), while for \( g_r > 0.013 \), the gradient term \( |\beta \vec{g}_i \chi(y_i)| \) dominates. In the high gradient regime, we observe the solid phase and this threshold therefore marks the crossover from the breaking to the solid regime.

\subsection{Low Gradient, Fluid and Breaking Regime $(g_r < g_{bs})$}
\label{SI.4.2}
As we reduce the gradient below this breaking-solid threshold, \( g_r < g_{bs} \), 
the next critical point occurs when the self-persistence term and the gradient term balance:
\[
\hat{v}_i(t - \Delta t) \approx \beta \vec{g}_i \chi(y_i).
\]

Again, solving for the critical gradient \( g_{fb} \), we find:
\[
g_{fb} = \frac{40}{19750} \approx 0.002.
\]

Thus, we identify three regions based on the gradient:
\begin{itemize}
  \item \( g_r < g_{fb} \): In this regime, motion is purely dominated by flocking behavior. \( \hat{v}_i(t - \Delta t),  \alpha \hat{V}  > \beta \vec{g}_i \chi(y_i) \)
  \item \( g_{fb} < g_r < g_{bs} \): In this intermediate regime, the gradient is able to compete with single-persistence effects but is not strong enough to overcome the overall collective behavior. This could explain the breakup observed in this domain.  \( \hat{v}_i(t - \Delta t) < \beta \vec{g}_i \chi(y_i) <  \alpha \hat{V} \).
  \item \( g_r > g_{bs} \): Here, the gradient dominates the overall dynamics, driving the system into a gradient-dominated, solid regime. \( \hat{v}_i(t - \Delta t),  \alpha \hat{V}  < \beta \vec{g}_i \chi(y_i) \).
\end{itemize}

These thresholds, with \( g_{fb} = 0.002 \) and \( g_{bs} = 0.013 \), define the transitions between different regimes of particle motion driven by varying influences of self-propulsion, alignment and gradient effects. 



\section{Analysis of Cluster Properties as a Function of Concentration and Gradient}
\label{SI.5}
\begin{figure}[ht]
  \centering
  \includegraphics[width=\linewidth]{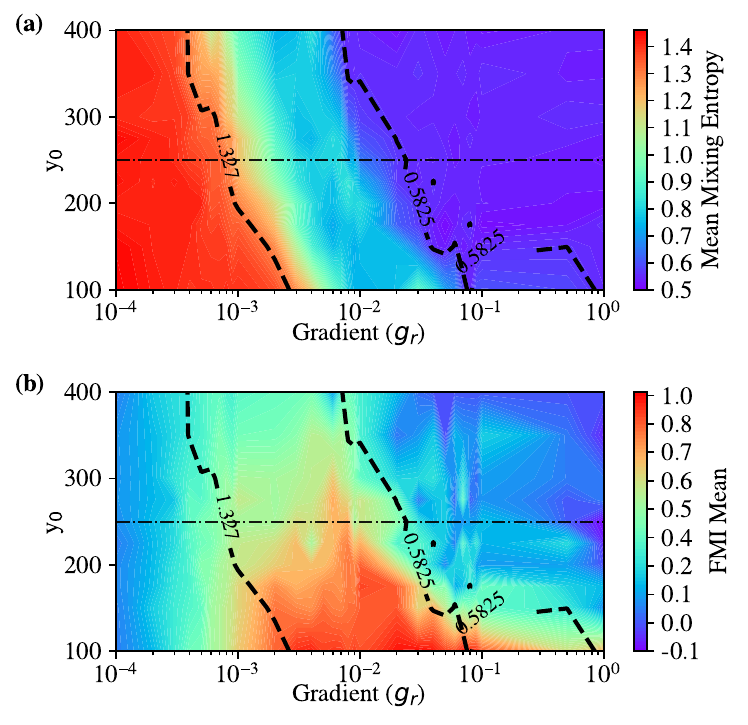}
  \caption{\textbf{\bf Dependence of Mixing Entropy and Cluster Migration on Concentration and Gradient:} (a) Mean mixing entropy as a function of gradient ($g_r$) and local concentration ($y_0$), illustrating regions of high and low mixing. (b) FMI showing the response of the cluster to external gradients as a function of COM concentration. The dashed contours indicate 10\% of maximum and minimum of entropy.
}
\label{FIG.SI1}
\end{figure}

In this section, we present additional analysis exploring the dynamical properties of the system (see Fig.~\ref{FIG.SI1}), for cluster size ($N=37$). Specifically, we investigate the mixing entropy and cluster response to external gradients as a function of the concentration at a position corresponding to the cluster's center of mass. The concentration at this point can be changed by changing the value of the parameter $y_0$ (Eq.~\ref{cy}). This analysis complements our primary results, providing insights into how the overall concentration influences cluster behavior in the presence of external gradients. The accompanying plots illustrate these dependencies, highlighting variations in mixing entropy and cluster response across different concentrations.\\


\section{FMI as a Function of Size}
\label{SI.6}

\begin{figure}[ht]
  \centering
  \includegraphics[width=\linewidth]{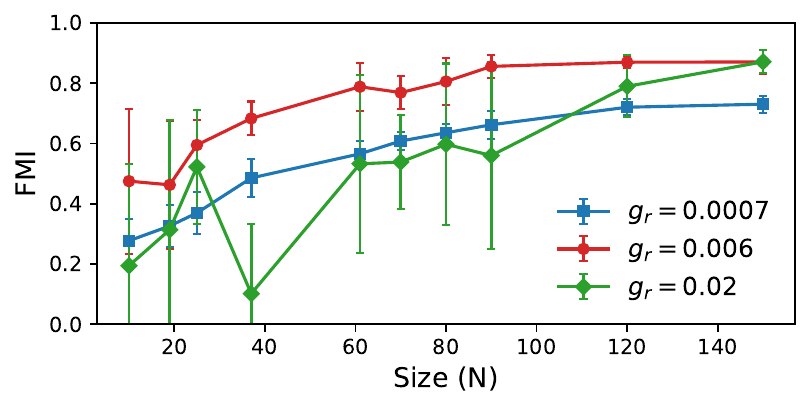}
  \caption{{\bf FMI value as a function of size (N)} for three different gradient values ($g_r = 0.0007$, $g_r = 0.006$, $g_r = 0.02$). The FMI variation for $g_r = 0.006$ (breaking regime) follows a trend similar to FMI observed in the Supplementary Figure~SI2($D^{ii}$) of \cite{malet-engraCollectiveCellMotility2015}, indicating that this gradient falls within the breaking regime.}
  \label{FIG.SI2}
\end{figure}

\section{Correlation of the Unit Active Propulsion}
\label{SI.7}
\begin{figure}[ht]
  \centering
  \includegraphics[width=0.98\linewidth]{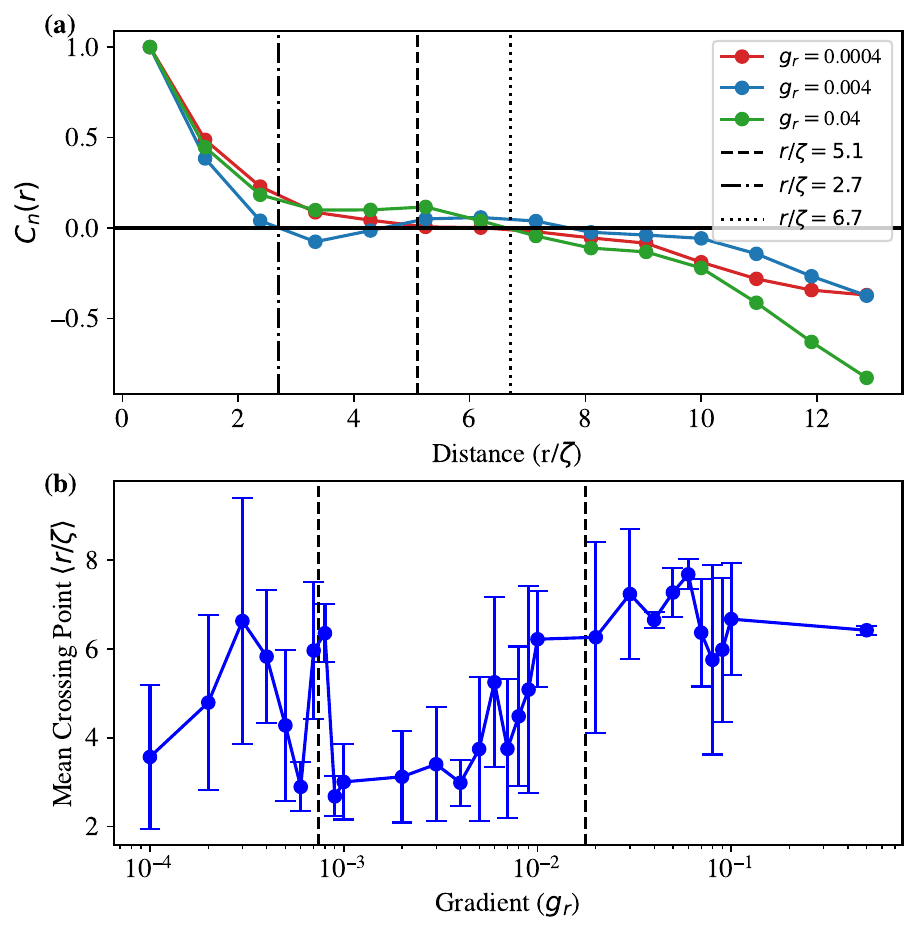}
  \caption{(a) \textbf{\bf Unit active propulsion correlation $C_n(r)$ as a function of scaled distance $r/\zeta$} for a fixed circular cluster of $N = 150$ in different regimes: fluid, breaking, and solid ($g_r =0.0004$, $g_r =0.004$ and $g_r =0.04$ respectively). The zero-crossing point of the correlation function is highest in the fluid regime, decreases in the breaking regime, and increases again in the solid regime, indicating a non-monotonic change in correlation length across these states. (b) {\bf Mean zero-crossing distance of the unit active propulsion correlation as a function of gradient values.} The trend follows an error function, highlighting the systematic variation of correlation length with gradient changes. This suggests a non-trivial dependence of structural stability on the imposed gradient.
}
\label{FIG.SI3}
\end{figure}

In our simulation, to investigate the reasons behind the onset of instabilities, we examined the correlation of the unit vector of active velocity within a circular cluster of $N=150$ particles, whose positions remain fixed throughout the analysis.

The unit active propulsion correlation function is defined as:
\begin{equation} 
C_n(r) = \frac{\sum_{i, j} \hat{n}_i \cdot \hat{n}_j \delta(r - r_{ij})}{\sum_{i, j} \delta(r - r_{ij})} 
\end{equation} 
where, $\hat{n}_i$ is the active velocity unit vector of particle $i$, and $r_{ij} = |\mathbf{r}_i - \mathbf{r}_j|$ is the separation distance between particles $i$ and $j$ and $\zeta$ is the nominal particle diameter (see Appendix~\ref{SI.2}). $\delta(r - r_{ij})$ is a binning function ensuring only pairs at distance $r$ contribute.

We analyzed this unit active propulsion correlation function in three different regimes: fluid, breaking, and solid ($g_r =0.0004$, $g_r =0.004$ and $g_r =0.04$ respectively). Our results indicate that the correlation function crosses zero at a higher distance in the fluid regime as seen in Fig.~\ref{FIG.SI3}(a).
This observation suggests that the structural properties of the system evolve in a non-monotonic manner across the different phases. The fluid regime, characterized by high mobility and disorder, exhibits the largest correlation length, implying a long-range velocity alignment among the particles. As the system moves into the breaking regime, where structural instability sets in, the correlation length significantly decreases, indicating a loss of coherence in velocity alignment. In the solid regime, where the system stabilizes into a more ordered configuration, the correlation length increases again, though not necessarily reaching the same values as in the fluid phase.

To further quantify this behavior, we analyzed the zero-crossing distances for various gradient values and averaged them over time (Fig.~\ref{FIG.SI3}(b)). 
On average, we observe a decrease in the crossover distance as we transition from the fluid to the breaking regime, indicating that smaller unit active propulsion correlation domains and the system becomes unstable. However, as the system further evolves into the solid cluster, the crossover distance increases again, reaffirming that structural stability is regained in the solid phase. This systematic variation of correlation length with the gradient suggests a fundamental relationship between active velocity fluctuations and phase stability in the system.

\section{Definition of Radius of Gyration}
\label{SI.8}
Furthermore, we performed shape analysis and dynamical analysis of these clusters. To understand the shape dynamics of the cluster, we calculated the radius of gyration and the asymmetry from the gyration tensor. The gyration tensor is a \(2 \times 2\) matrix that describes the spatial distribution of particles in a cluster relative to their center of mass. It is computed as follows: 
\[
\mathbf{G} = \frac{1}{N} \sum_{i=1}^{N} 
\begin{pmatrix} 
(x_i - x_{cm})^2 & (x_i - x_{cm})(y_i - y_{cm}) \\ 
(x_i - x_{cm})(y_i - y_{cm}) & (y_i - y_{cm})^2 
\end{pmatrix}
\]
where \( (x_i, y_i) \) are the coordinates of the \( i \)-th particle, and \( (x_{cm}, y_{cm}) \) are the coordinates of the center of mass of the cluster. 

To characterize the shape of the cluster, we analyzed the eigenvalues and eigenvectors of the gyration tensor. The eigenvalues (\(\lambda_1, \lambda_2\)) represent the principal variances of the particle distribution along specific directions, while the corresponding eigenvectors indicate the principal axes of the shape. If \( \mathbf{v}_1 = (v_{1x}, v_{1y}) \) and \( \mathbf{v}_2 = (v_{2x}, v_{2y}) \) are the eigenvectors corresponding to the eigenvalues \( \lambda_1 \) and \( \lambda_2 \), respectively, then \( \mathbf{v}_1 \) defines the direction along which the cluster is most extended, while \( \mathbf{v}_2 \) represents the perpendicular direction. 





\section{Average Orientation Angle as a Function of Gradient}
\label{SI.9}
\begin{figure}[ht]
  \centering
  \includegraphics[width=\linewidth]{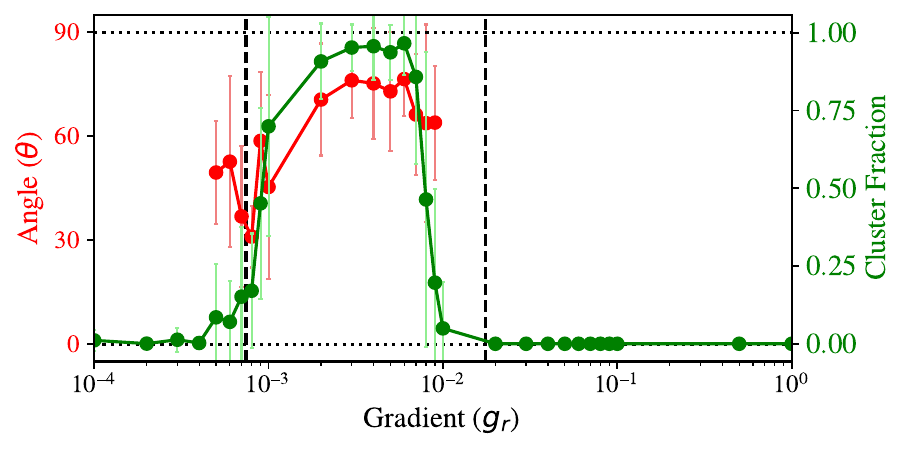}
  \caption{\textbf{Average orientation angle ($\theta$) (if the cluster fraction exceeds 5\%) as a function of gradient}, illustrating alignment changes with increasing gradient. The inset plot also includes the fraction of cluster with aspect ratio $> 2$, showing the relationship between orientation dynamics and the presence of elongated clusters across gradient regimes.
}
\label{FIG.SI4}
\end{figure}
Fig.~\ref{FIG.SI4} illustrates that the average orientation angle (\(\theta\)) varies as a function of the gradient, but only for clusters with significant elongation (AR \( > 2 \)). These plots also show the cluster fraction across gradient regimes, emphasizing how orientation dynamics are closely linked to the prevalence of elongated clusters in the breaking regime.

\section{Velocity of Clusters in the Direction of the Chemical Gradient}
\label{SI.10}

\begin{figure}[ht]
  \centering
  \includegraphics[width=\linewidth]{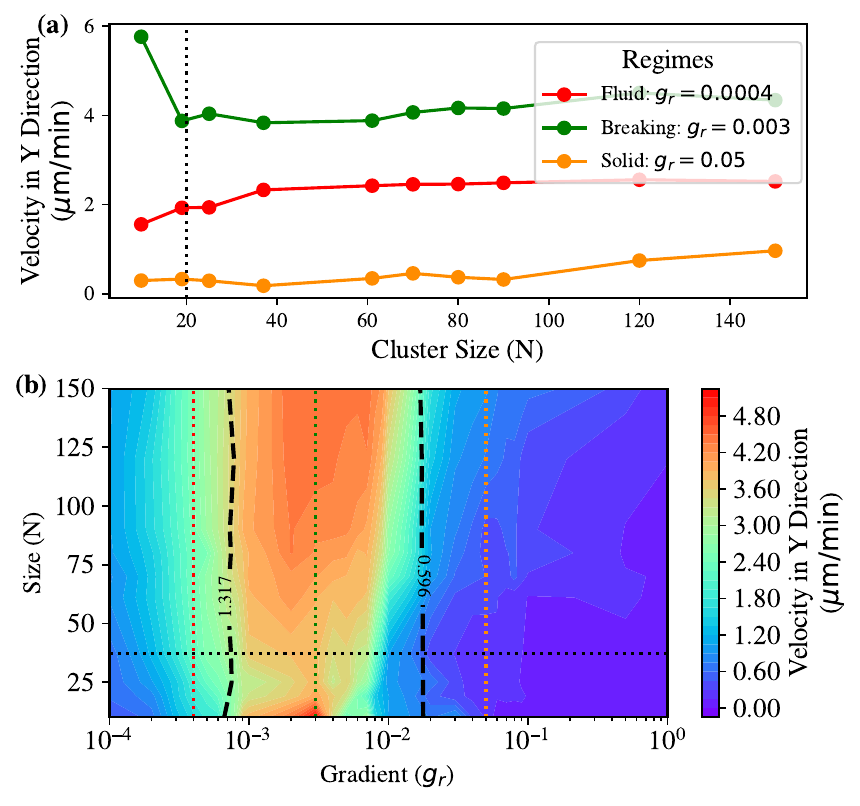}
  \caption{{\bf Cluster velocity in the Y direction as a function of cluster size and gradient strength.} (a) Y-direction velocity across the fluid, breaking, and solid regimes under varying gradient strengths ($g_r$). Dashed lines indicate the cluster size below which breakup does not occur in the breaking regime. The Y direction velocity is notably elevated in this regime. 
  (b) Heatmap of cluster velocity in the Y direction as a function of cluster size and gradient strengths, with dashed lines indicating transitions between regimes. The heatmap further reinforces that the response to the gradient is most pronounced in the breaking regime.}
  \label{FIG.SI5}
\end{figure}

The key focus here is the cluster's response to the gradient, characterized by its velocity in the direction of the gradient i.e. velocity in the Y direction. 
The resulting velocity in the Y direction is notably higher in the breaking regime (see \ref{FIG.SI5}(a,b)). 

\section*{Supplementary Videos (G-Drive)}
\label{SI.11}

Please find the videos available at \href{https://drive.google.com/drive/folders/1GNubL6CEZdfDJj28hi8LJJ9rQuB1Iznq?usp=sharing}{(Google Drive link)}.

\begin{enumerate}
  \item \textbf{Video M1.} Migration of T and B lymphocytes under a 0–100 ng/ml/mm CCL19 gradient.
  
  \item \textbf{Video M2.} Time-lapse of cluster migration under a 0–100 ng/ml/mm CCL19 gradient (see Fig.~\ref{FIG.main1}).

  \item \textbf{Video M3.} Fluid regime (simulation): dynamic rearrangement and intermixing of colored particles (see Fig.~\ref{FIG.main3}(e)).

  \item \textbf{Video M4.} Breaking regime (simulation): deformation and partial mixing preceding fragmentation (see Fig.~\ref{FIG.main3}(f)).

  \item \textbf{Video M5.} Solid regime (simulation): minimal rearrangement and preserved color pattern (see Fig.~\ref{FIG.main3}(g)).

  \item \textbf{Video M6.} Low-gradient regime (simulation): weak alignment between propulsion vectors ($\hat{n}$, purple arrows) and the chemical gradient ($\vec{g}$, black arrows) leads to uncoordinated, fluid-like motion (see Fig.~\ref{FIG.main4}(a)).

  \item \textbf{Video M7.} Breaking regime (simulation): fluctuating alignment of propulsion vectors leads to cluster deformation and fragmentation.

  \item \textbf{Video M8.} High-gradient regime (simulation): strong alignment of propulsion vectors stabilizes the cluster into a compact structure (see Fig.~\ref{FIG.main4}(b)).

  \item \textbf{Video M9.} Cell migration under a 0–50 ng/ml/mm CCL19 gradient. 
    
  \item \textbf{Video M10.} Fragmentation of a medium cluster under a 0–100 ng/ml/mm CCL19 gradient (see Fig.~\ref{FIG.main5}(a)).


  \item \textbf{Video M11.} Simulation of medium cluster breakup, consistent with Videos M10 (see Fig.~\ref{FIG.main5}(b)).

  \item \textbf{Video M12.} Simulation showing trajectories of selected particles in a migrating cluster (see Fig.~\ref{FIG.main6}(a)) in breaking regime.

  \item \textbf{Video M13.} Large cluster migration and elongation under a 0–100 ng/ml/mm CCL19 gradient (see Fig.~\ref{FIG.main7}(a)).


  \item \textbf{Video M14.} Simulation of large cluster fragmentation, consistent with Videos M13 (see Fig.~\ref{FIG.main7}(b)).
\end{enumerate}

\begin{acknowledgments}
This work was supported by the National Science Foundation (NSF-DMS-1616926 to A.G.) and NSF-CREST: Center for Cellular and Bio-molecular Machines at UC Merced (NSF-HRD-1547848 and EES-2112675 to A.G.). A.G. and M.S. also acknowledge partial support from the NSF Center for Engineering Mechanobiology grant CMMI-154857 and computing time on the Multi-Environment Computer for Exploration and Discovery (MERCED) cluster at UC Merced (NSF-ACI-1429783). This research also benefited from the Center for Living Systems (NSF grant no.  2317138)
N.S.G. is supported by the Lee and William Abramowitz Professorial Chair of Biophysics (Weizmann Institute) with additional support from a Royal Society Wolfson Visiting Fellowship.
Work in G.S.'s \ laboratory was supported by ERC-Synergy (Grant\# 801 101071470), AIRC-IG (Grant\#22821), AIRC 5x1000 (\#22759), the Italian Ministry of University and Research (PRIN202223GSCIT\_01/ G53D23002570006/20229RM8A\_001; COMBINE/ G53D23007040001/ P2022RH4HH002; PNRR\_CN3RNA\_SPOKE/ G43C22001320007).


\end{acknowledgments}

\bibliography{rsc, references, References_overleaf, toCheck, MyLibrary} 

\end{document}